\documentclass[twocolumn,showpacs,amsmath,amssymb,prd]{revtex4}

\usepackage{graphicx}
\usepackage{dcolumn}
\usepackage{bm}
\usepackage[varg]{txfonts}

\begin{document}

\title{Resumming Cosmological Perturbations via the Lagrangian
  Picture:\\ One-loop Results in Real Space and in Redshift Space }

\author{Takahiko Matsubara}
\email{taka@a.phys.nagoya-u.ac.jp}
\affiliation{%
Department of Physics, Nagoya University,
Chikusa, Nagoya, 464-8602, Japan
}%

\date{\today}

\begin{abstract}
    We develop a new approach to study the nonlinear evolution in the
    large-scale structure of the Universe both in real space and in
    redshift space, extending the standard perturbation theory of
    gravitational instability. Infinite series of terms in standard
    Eulerian perturbation theory are resummed as a result of our
    starting from a Lagrangian description of perturbations. Delicate
    nonlinear effects on scales of the baryon acoustic oscillations
    are more accurately described by our method than the standard one.
    Our approach differs from other resummation techniques recently
    proposed, such as the renormalized perturbation theory, etc., in
    that we use simple techniques and thus resulting equations are
    undemanding to evaluate, and in that our approach is capable of
    quantifying the nonlinear effects in redshift space. The power
    spectrum and correlation function of our approach are in good
    agreement with numerical simulations in literature on scales of
    baryon acoustic oscillations. Especially, nonlinear effects on the
    baryon acoustic peak of the correlation function are accurately
    described both in real space and in redshift space. Our approach
    provides a unique opportunity to analytically investigate the
    nonlinear effects on baryon acoustic scales in observable redshift
    space, which is requisite in constraining the nature of dark
    energy, the curvature of the Universe, etc., by redshift surveys.
\end{abstract}

\maketitle

\section{\label{sec:intro}
Introduction
}

Recent progress in cosmology is largely stimulated by precise
measurements of the Universe. The redshift survey of galaxies is one
of the most important methods in distinguishing cosmological models.
The baryon acoustic oscillations (BAOs) in the early universe
\cite{BAOtheory}, which were first detected in the cosmic microwave
background (CMB) anisotropy power spectrum \cite{CMBobs}, were also
detected in the large-scale structure probed by redshift surveys
\cite{BAOLSS}. The expansion history and curvature of the Universe can
be geometrically measured by galaxy clustering in redshift space
\cite{AP79, GeomTest}, and the BAO signature in the large-scale
structure yields a standard ruler \cite{BAOSR} for this purpose.
Therefore, accurately measuring the BAO scales in a statistical
quantity such as the power spectrum or correlation function of
galaxies, one can constrain the structure of the Universe, such as the
nature of dark energy \cite{BAODE}.

To utilize the BAOs as a standard ruler in the Universe, a
theoretically precise description of the BAOs is of the utmost
importance. The physics of the BAOs at sufficiently large redshift is
well described by linear theory. Only the amplitude of the spatial
pattern of clustering is increased, and the characteristic scale of
the BAOs is unaltered in this regime. However, the nonlinear dynamics
play an important role in a redshift range where we can measure the
large-scale structure. Even though the BAO scales $\simeq
100\,h^{-1}{\rm Mpc}$ are large, the nonlinearity deforms the BAO
pattern in the power spectrum and correlation function \cite{SE05,
  Spr05, BAONL, BAONLRD, ang07, ESW07}. Nonlinear deformations of the
power spectrum are described by the standard perturbation theory (SPT)
\cite{jus81, vish83, Fry84, Goroff86, MSS, JB94, BCGS02} of higher
orders as long as the density fluctuations on BAO scales are still in
the linear regime at sufficiently large redshift \cite{MWP99, JK06}.
The last condition is not necessarily satisfied by realistic surveys.
Moreover, when the cold dark matter (CDM) is present, the higher-order
SPT does not give any sensible prediction for the correlation function
because unphysical behavior in small wavelength limit prevents Fourier
transform from converging.

Recently, several other approaches beyond the framework of SPT have
been proposed \cite{Sco01, NewPT, CS06a, CS06b, Val07, CS07, RGPT,
  TH07}. Among others, Scoccimarro's reformulation of fluid equations
using the propagator, the vertex, and a source \cite{Sco01} provides a
way to use standard tools of field theory, yielding the renormalized
perturbation theory (RPT) \cite{CS06a, CS06b, CS07}, the large-$N$
expansions \cite{Val07}, the renormalization group approach
\cite{RGPT}, and the closure theory \cite{TH07} in the context of
gravitational instability theory. In these newly developed approaches,
the standard perturbative expansion is reorganized and partially
resummed in various ways. It is still difficult to exactly calculate
the power spectrum by these approaches, and different levels of
approximations and ansatz are employed. The predictions of the power
spectrum calculated by the 2-loop RPT and renormalization group method
are shown to agree with measurements of the $N$-body simulations on
BAO scales even at low redshift where the 1-loop SPT breaks down
\cite{CS07, RGPT}. It is still nontrivial why these approaches work
well despite their approximate treatments and use of ansatz.

So far the predictions of the above nonlinear theories, such as SPT,
RPT, and their variants, have been confined in real space. Since the
large-scale structure is actually measured in redshift space, it is
mandatory to make predictions in redshift space for applying the
nonlinear theories to realworld observations. The redshift-space
distortion effects in the linear regime are analytically given by
first-order SPT (i.e., linear theory) \cite{Kaiser87, Ham92,
  GeomTest}. However, investigations of nonlinear redshift-space
distortions by higher-order SPT are quite limited in literature
\cite{HMV98,SCF99}, apart from numerical simulations and nonlinear
modelings \cite{MWP99, BAONLRD, ang07, Sco04, SE05, ESW07}.

In this paper, we consider resummations of the perturbative expansion
by a different approach from RPT and its variants. Instead of relying
on Scoccimarro's reformulation, we start from a Lagrangian picture of
density perturbations. Why do we have to add yet another approach even
though there are already enough? There are a number of reasons. First,
our approach is simpler than other resummation methods. While it is
straightforward to calculate the power spectrum in our approach, other
resummation techniques are so complicated that approximate treatments
and/or ansatz are necessary in deriving the final power spectrum.
Second, as a result of that, our analytic expression of the resulting
power spectrum is no more complicated than the SPT expression, and
numerical evaluations are fast performed. Third, and most importantly,
our approach is capable of evaluating the nonlinear power spectrum
{\em in redshift space}, which is beyond the scope of other
resummation methods in literature. Fourth, unlike in SPT, the power
spectrum in our approach can be Fourier transformed to predict the
nonlinear effects in the correlation function.

An important ingredient of our approach is the perturbation theory in
Lagrangian space, which is known as the Lagrangian perturbation theory
(LPT) \cite{LPT}. The first-order LPT corresponds to the classic
Zel'dovich approximation \cite{Zel70}. Since the observable power
spectrum is given in Eulerian space, predicting the power spectrum by
LPT is not a trivial task. We define a method of calculating the power
spectrum from the LPT, partially expanding the Lagrangian variables in
Eulerian space. The resulting expression contains an infinite series
of SPT terms; that is, the SPT terms are automatically resummed in our
formulation. The underlying motivation of this formulation is shared
with RPT. Readers are recommended to go through an excellent
introduction in Ref.~\cite{CS06a} using the Zel'dovich approximation,
which equally applies to our formulation since our formalism and RPT
become equivalent in a case of Zel'dovich approximation.

The rest of this paper is organized as follows. In
Sec.~\ref{sec:Resum}, our basic formalism is described. A brief review
of LPT is also included. In Sec.~\ref{sec:RealSpace}, analytic results
of the power spectrum and correlation function in real space are
presented and compared with simulation results in the literature. The
corresponding results in redshift space are derived in
Sec.~\ref{sec:RedshiftSpace}. The conclusions are given in
Sec.~\ref{sec:concl}.

\section{\label{sec:Resum} Resumming Perturbations via
the Lagrangian Picture}

\subsection{\label{subsec:LagPS}
The power spectrum from the Lagrangian variable
}

In a Lagrangian picture, a fundamental variable to represent
perturbations is a displacement field $\bm{\Psi}(\bm{q},t)$, which maps
a fluid element from initial Lagrangian coordinates $\bm{q}$ to the
Eulerian coordinates $\bm{x}$ at a time $t$:
\begin{equation}
  \bm{x}(\bm{q},t) = \bm{q} + \bm{\Psi}(\bm{q},t).
\label{eq:1-1}
\end{equation}
We will drop the argument $t$ in the following for notational
simplicity, and readers should keep in mind that the dynamical
variables always depend on time.

Assuming the initial density field is sufficiently uniform, the
Eulerian density field $\rho(\bm{x})$ satisfies the continuity
relation, $\rho(\bm{x}) d^3x = \bar{\rho}d^3q$, where $\bar{\rho}$ is
the mean density in comoving coordinates. This relation is equivalent
to the following equation:
\begin{equation}
  \delta(\bm{x}) = \int d^3q\,
  \delta^3\left[\bm{x} - \bm{q} - \bm{\Psi}(\bm{q})\right] - 1,
\label{eq:1-2}
\end{equation}
where $\delta = \rho/\bar{\rho}-1$ is the density contrast. The power
spectrum $P(\bm{k})$ is defined by a relation
\begin{equation}
  \left\langle\tilde{\delta}(\bm{k})\tilde{\delta}(\bm{k}')\right\rangle
  = (2\pi)^3 \delta^3(\bm{k} + \bm{k}') P(\bm{k}),
\label{eq:1-3}
\end{equation}
where 
\begin{equation}
  \tilde{\delta}(\bm{k})
  = \int d^3x\, e^{-i\bm{k}\cdot\bm{x}} \delta(\bm{x})
\label{eq:1-4}
\end{equation}
is the Fourier transform of the density contrast. The power spectrum
in real space depends only on the magnitude of the wave vector,
$k=|\bm{k}|$. In redshift space, however, a directional dependence
emerges. Using Eqs.~(\ref{eq:1-2})--(\ref{eq:1-4}), the power spectrum
can be expressed by a displacement field as \cite{TaylorPS}
\begin{equation}
  P(\bm{k}) = \int d^3q\, e^{-i\bm{k}\cdot\bm{q}}
  \left(
      \left\langle
          e^{-i\bm{k}\cdot
            [\bm{\Psi}(\bm{q}_1) - \bm{\Psi}(\bm{q}_2)]}
      \right\rangle - 1
  \right),
\label{eq:1-5}
\end{equation}
where $\bm{q} = \bm{q}_1 - \bm{q}_2$. Because of the homogeneity, the
ensemble average in the integrand of the right-hand side (RHS) only
depends on the separation $\bm{q}$. The expression of
Eq.~(\ref{eq:1-5}) is a general relation between the power spectrum
and displacement fields.

We use the cumulant expansion theorem \cite{Ma85}
\begin{equation}
  \langle e^{-iX} \rangle
  = \exp\left[\sum_{N=1}^\infty \frac{(-i)^N}{N!} \langle X^N
      \rangle_{\rm c} \right],
\label{eq:1-6}
\end{equation}
in Eq.~(\ref{eq:1-5}), where $\langle X^N \rangle_{\rm c}$ denotes a
cumulant of a random variable $X$ \cite{BCGS02}. In the resulting
equation, we have a following quantity:
\begin{align}
&
    \left\langle
        \left\{
            \bm{k}\cdot\left[
                \bm{\Psi}(\bm{q}_1) - \bm{\Psi}(\bm{q}_2)
            \right]
        \right\}^N
    \right\rangle_{\rm c}
\nonumber\\
& \quad
    = [1 + (-1)^N]
    \left\langle[\bm{k}\cdot\Psi(\bm{0})]^N\right\rangle_{\rm c}
\nonumber\\
& \quad
    + \sum_{j=1}^{N-1}
    (-1)^{N-j} \left(\begin{array}{c}N\\j\end{array}\right)
    \left\langle
        [\bm{k}\cdot\Psi(\bm{q}_1)]^j [\bm{k}\cdot\Psi(\bm{q}_2)]^{N-j}
    \right\rangle_{\rm c},
\label{eq:1-7}
\end{align}
where the binomial theorem is applied. In this equation, only $N\geq
2$ cases are survived, because $\langle\bm{\Psi}\rangle = 0$ for
parity symmetry. Accordingly, we obtain a general expression,
\begin{align}
    & P(\bm{k}) =
    \exp
    \left[-2\sum_{n=1}^\infty
        \frac{k_{i_1}\cdots k_{i_{2n}}}{(2n)!}
        A^{(2n)}_{i_1 \cdots i_{2n}}
    \right]
    \int d^3q\, e^{-i\bm{k}\cdot\bm{q}}
\nonumber\\
    & \qquad\quad \times
    \left\{
        \exp\left[ \sum_{N=2}^\infty
            \frac{k_{i_1}\cdots k_{i_N}}{N!}
            B^{(N)}_{i_1 \cdots i_N}(\bm{q})
        \right] - 1
    \right\},
\label{eq:1-8}
\end{align}
where summation over repeated indices is assumed as usual, and
\begin{align}
    & A^{(2n)}_{i_1 \cdots i_{2n}} = 
    (-1)^{n-1} \left\langle\Psi_{i_1}(\bm{0})\cdots
        \Psi_{i_{2n}}(\bm{0})\right\rangle_{\rm c},
\label{eq:1-9a}\\
    & B^{(N)}_{i_1 \cdots i_N}(\bm{q}) = 
    i^N \sum_{j=1}^{N-1}
    (-1)^j \left(\begin{array}{c}N\\j\end{array}\right)
\nonumber\\
    & \quad \times
    \left\langle
        \Psi_{(i_1}(\bm{q}_1)\cdots
        \Psi_{i_{j}}(\bm{q}_1)
        \Psi_{i_{j+1}}(\bm{q}_2)\cdots
        \Psi_{i_N)}(\bm{q}_2)
    \right\rangle_{\rm c},
\label{eq:1-9b}
\end{align}
are defined. In the RHS of Eq.~(\ref{eq:1-9b}), all the $N$ indices
are symmetrized over.

The quantity $A^{(N)}$ of Eq.~(\ref{eq:1-9a}) is given by a cumulant
of a displacement vector at a single position. The quantity $B^{(N)}$
of Eq.~(\ref{eq:1-9b}) is given by cumulants among two displacement
vectors which are separated by a Lagrangian distance $|\bm{q}|$.
Correlations among components of separated displacement vectors are
expected to be small when the separations are large. When we are
interested in clustering on large scales, the magnitude of wave vector
$k$ is small, and large separations of $|\bm{q}|$ in
Eq.~(\ref{eq:1-5}) efficiently contribute to the integral. Therefore,
we expect the cumulants of $B^{(N)}$ are smaller than the cumulants of
$A^{(N)}$ on large scales.

With the above consideration, we propose to {\em expand only the
  $B^{(N)}$ terms in Eq.~(\ref{eq:1-8}) as small quantities and keep
  the $A^{(N)}$ terms in the exponent}. If the $A^{(N)}$ terms are
expanded as well, this equation turns out to give equivalent results
of SPT, which is formulated in Eulerian space, as shown below.
Therefore, the exponential prefactor contains infinitely higher-order
perturbations in terms of SPT. Thus our method provides a way to resum
the infinite series of perturbations in the SPT. This point is our
first key technique of this paper. The second technique is to evaluate
the cumulants of the displacement field by the LPT, which is addressed
in the following. Before that, it is convenient to represent the
cumulants of the displacement field in Fourier space.

\subsection{\label{subsec:psdis}
Polyspectra of the displacement field
}

To evaluate Eq.~(\ref{eq:1-8}), it is useful to define the polyspectra
$C_{i_1 \cdots i_N}$ of the displacement field by
\begin{align}
&
  \left\langle
    \tilde{\Psi}_{i_1} (\bm{p}_1) \cdots
    \tilde{\Psi}_{i_N} (\bm{p}_N)
  \right\rangle_{\rm c}
\nonumber\\
& \quad =
 (2\pi)^3 \delta^3(\bm{p}_1 + \cdots + \bm{p}_N)
   (-i)^{N-2}  C_{i_1 \cdots i_N}(\bm{p}_1,\ldots,\bm{p}_N),
\label{eq:1-10}
\end{align}
where
\begin{equation}
  \tilde{\Psi}_i (\bm{p}) = 
  \int d^3q\,e^{-i\bm{p}\cdot\bm{q}}
  {\Psi}_i (\bm{q})
\label{eq:1-11}
\end{equation}
is the Fourier transform of the displacement field. The relation
$\bm{p}_1 + \cdots + \bm{p}_N = \bm{0}$ is always satisfied because of
the translational invariance. For $N=2$, we employ a notation,
$C_{ij}(\bm{p}) = C_{ij}(\bm{p},-\bm{p})$, for simplicity. The factors
$(-i)^{N-2}$ in the RHS of Eq.~(\ref{eq:1-10}) are there to ensure
that the polyspectra $C_{i_1\cdots}$ are real numbers. In fact, one
easily finds the complex conjugate satisfies
\begin{equation}
  \left\langle
    \tilde{\Psi}_{i_1}(\bm{p}_1) \cdots
    \tilde{\Psi}_{i_N}(\bm{p}_N)
  \right\rangle_{\rm c}^*
  = (-1)^N \left\langle
    \tilde{\Psi}_{i_1}(\bm{p}_1) \cdots
    \tilde{\Psi}_{i_N}(\bm{p}_N)
  \right\rangle_{\rm c},
\label{eq:1-12}
\end{equation}
which is an equivalent condition that the polyspectra are real
numbers, $C_{i_1\cdots i_N}^* = C_{i_1\cdots i_N}$. It is also
important to note the polyspectra of the displacement field satisfy
a parity relation,
\begin{equation}
  C_{i_1 \cdots i_N}(-\bm{p}_1,\ldots,-\bm{p}_N)
  = (-1)^N C_{i_1 \cdots i_N}(\bm{p}_1,\ldots,\bm{p}_N),
\label{eq:1-13}
\end{equation}
which is explicitly shown by Eqs.~(\ref{eq:1-10}) and (\ref{eq:1-11}).

Using the polyspectra defined above, Eqs.~(\ref{eq:1-9a}) and
(\ref{eq:1-9b}) reduce to 
\begin{align}
    & A^{(2n)}_{i_1 \cdots i_{N}} = 
    \int \frac{d^3p_1}{(2\pi)^3} \cdots \frac{d^3p_{2n}}{(2\pi)^3}
    \delta^3(\bm{p}_1 + \cdots + \bm{p}_{2n})
\nonumber\\
& \qquad\qquad\qquad\qquad\qquad\quad \times
    C_{i_1\cdots i_{2n}}(\bm{p}_1,\ldots,\bm{p}_{2n}),
\label{eq:1-14a}\\
    & B^{(N)}_{i_1 \cdots i_N}(\bm{q}) = 
    \sum_{j=1}^{N-1}
    (-1)^{j-1} \left(\begin{array}{c}N\\j\end{array}\right)
\nonumber\\
    & \qquad\qquad \times
    \int \frac{d^3p_1}{(2\pi)^3} \cdots \frac{d^3p_{N}}{(2\pi)^3}
    \delta^3(\bm{p}_1 + \cdots + \bm{p}_{N})
\nonumber\\
& \qquad\qquad\qquad \times
    e^{i(\bm{p}_1 + \cdots + \bm{p}_j)\cdot\bm{q}}
    C_{i_1\cdots i_{N}}(\bm{p}_1,\ldots,\bm{p}_{N}).
\label{eq:1-14b}
\end{align}


\subsection{\label{subsec:LPT}
The Lagrangian perturbation theory
}

For a Newtonian pressureless self-gravitating fluid embedded in an
expanding universe, the displacement field is governed by the equation
of motion,
\begin{equation}
  \frac{d^2\bm{\Psi}}{dt^2} + 2 H \frac{d\bm{\Psi}}{dt}
  = - \bm{\nabla}_x\phi[\bm{q} + \bm{\Psi}(\bm{q})],
\label{eq:1-16}
\end{equation}
where $H = \dot{a}/a$ is the time-dependent Hubble parameter, and
$\bm{\nabla}_x$ is a derivative with respect to the Eulerian coordinate,
$\bm{x} = \bm{q} + \bm{\Psi}(\bm{q})$. The gravitational potential
$\phi$ is determined by the Poisson equation,
\begin{equation}
  \bm{\nabla}_x^2 \phi(\bm{x})
  = 4\pi G \bar{\rho} a^2 \delta(\bm{x}).
\label{eq:1-17}
\end{equation}
where the density contrast is related to the displacement field by
Eq.~(\ref{eq:1-2}) or, equivalently,
\begin{equation}
  \delta(\bm{x})
  = \left\{
      {\rm det}\left[\delta_{ij} + \Psi_{i,j}(\bm{q})\right]
    \right\}^{-1} -1. 
\label{eq:1-18}
\end{equation}

We use the LPT \cite{LPT} to evaluate the polyspectra of the displacement
field. In LPT, the displacement field is expanded by a perturbative
series:
\begin{equation}
  \bm{\Psi} = \bm{\Psi}^{(1)} +  \bm{\Psi}^{(2)} +  \bm{\Psi}^{(3)} + 
  \cdots,
\label{eq:1-19}
\end{equation}
where $\bm{\Psi}^{(N)}$ has the order of $(\bm{\Psi}^{(1)})^N$, which
is considered a small quantity (properly speaking, dimensionless
quantity $\partial_i\Psi_j$ is considered to be small). It is quite
common in the LPT that the velocity field is assumed to be
irrotational,
\begin{equation}
  \bm{\nabla}_x\times \bm{v} = \bm{0}.
\label{eq:1-20}
\end{equation}
This condition is compatible with the dynamical Eq.~(\ref{eq:1-16})
and restricts one to considering a subclass of general solutions. The
main motivation of this restriction is that the vortical motions
linearly decay in Eulerian perturbation theory \cite{BCGS02} and thus
are negligible in the quasilinear regime. We assume this condition
throughout this paper. The longitudinal part of Eq.~(\ref{eq:1-16}) is
an independent equation to be solved perturbatively in LPT; the same
is true for Eqs.~(\ref{eq:1-17}), (\ref{eq:1-18}), and
(\ref{eq:1-20}). Each perturbative term of the displacement field in
Fourier space is generally represented as
\begin{align}
& \tilde{\bm{\Psi}}^{(n)}(\bm{p}) =
  \frac{i D^n}{n!}
  \int \frac{d^3p_1}{(2\pi)^3} \cdots \frac{d^3p_n}{(2\pi)^3}
  (2\pi)^3 \delta^3\left(\sum_{j=1}^n \bm{p}_j - \bm{p}\right)
\nonumber\\
& \qquad\qquad\qquad\quad \times
  \bm{L}^{(n)}(\bm{p}_1,\ldots,\bm{p}_n)
  \delta_0(\bm{p}_1) \cdots \delta_0(\bm{p}_n),
\label{eq:1-21}
\end{align}
where $\delta_0 = \tilde{\delta}^{(1)}(t=t_0)$ is the linear density
contrast at the present time $t_0$, $D(t)$ is a linear growth rate
normalized by $D(t_0) = 1$. The imaginary unit is there to ensure that
the perturbative kernels $\bm{L}^{(n)}$ are real vectors. This is
straightforward to show by considering the parity transformation of
both sides of Eq.~(\ref{eq:1-21}). The perturbative kernels do not
depend on time in an Einstein--de~Sitter model. In general cosmology,
they weakly depend on time and cosmological parameters \cite{BCHJ95}.
It is still a good approximation that the perturbative kernels in
arbitrary cosmology are replaced by those in the Einstein--de~Sitter
model \cite{BCGS02}. We adopt this approximation throughout this paper
for simplicity, although including such dependence is straightforward.
The perturbative kernels in LPT up to third order are given by
\cite{Cat95}
\begin{align}
& \bm{L}^{(1)}(\bm{p}_1) = \frac{\bm{k}}{k^2}
\label{eq:1-22a}\\
& \bm{L}^{(2)}(\bm{p}_1,\bm{p}_2)
  =\frac37 \frac{\bm{k}}{k^2}
  \left[1 - \left(\frac{\bm{p}_1 \cdot \bm{p}_2}{p_1 p_2}\right)^2\right]
\label{eq:1-22b}\\
& \bm{L}^{\rm (3a)}(\bm{p}_1,\bm{p}_2,\bm{p}_3)
\nonumber\\
& \quad
  = \frac57 \frac{\bm{k}}{k^2}
  \left[1 - \left(\frac{\bm{p}_1 \cdot \bm{p}_2}{p_1 p_2}\right)^2\right]
  \left\{1 - \left[\frac{(\bm{p}_1 + \bm{p}_2) \cdot \bm{p}_3}
          {|\bm{p}_1 + \bm{p}_2| p_3}\right]^2\right\}
\nonumber\\
& \qquad
  - \frac13 \frac{\bm{k}}{k^2}
  \left[
      1 - 3\left(\frac{\bm{p}_1 \cdot \bm{p}_2}{p_1 p_2}\right)^2
  \right.
\nonumber\\
& \qquad\qquad\qquad\qquad
  \left.
      +\, 2 \frac{(\bm{p}_1 \cdot \bm{p}_2)(\bm{p}_2 \cdot \bm{p}_3)
        (\bm{p}_3 \cdot \bm{p}_1)}{p_1^2 p_2^2 p_3^2}
  \right]
\nonumber\\
& \qquad
  + \bm{k}\times \bm{T}(\bm{p}_1,\bm{p}_2,\bm{p}_3),
\label{eq:1-22c}
\end{align}
where $\bm{k} = \bm{p}_1 + \cdots + \bm{p}_n$ for $\bm{L}^{(n)}$, and
a vector $\bm{T}$ represents a transverse part whose expression is not
needed in the following application. It is useful to symmetrize the
kernel $\bm{L}^{\rm (3a)}$ in terms of their arguments:
\begin{equation}
  \bm{L}^{(3)}(\bm{p}_1,\bm{p}_2,\bm{p}_3) =
  \frac13
  \left[\bm{L}^{\rm (3a)}(\bm{p}_1,\bm{p}_2,\bm{p}_3) + {\rm perm.}\right].
\label{eq:1-23}
\end{equation}

\subsection{\label{subsec:OneLoopPS}
One-loop approximation to the resummed power spectrum
}

\begin{figure}
\includegraphics[width=20pc]{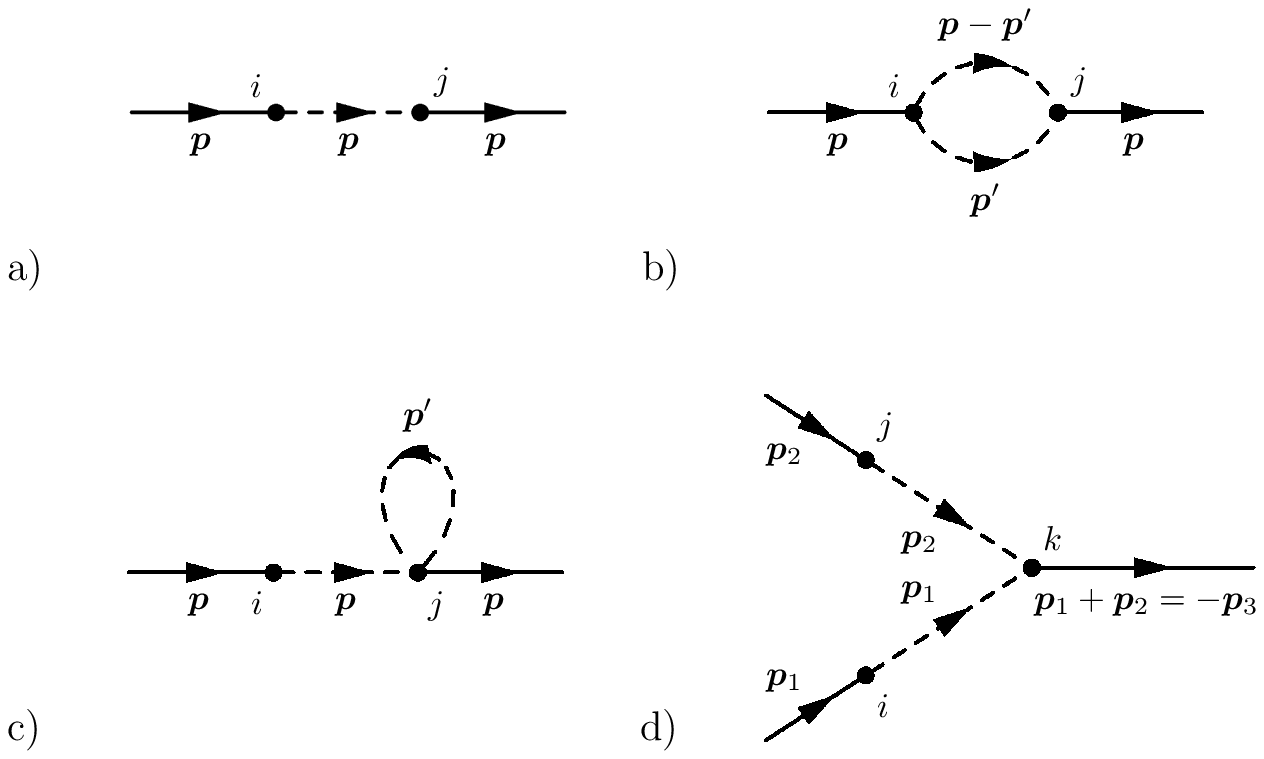}
\caption{\label{fig:polydiag} Diagrammatic representations of
  $C^{(11)}_{ij}(\bm{p})$ (a), $C^{(22)}_{ij}(\bm{p})$ (b),
  $C^{(13)}_{ij}(\bm{p})$ (c), and
  $C^{(112)}_{ijk}(\bm{p}_1,\bm{p}_2,\bm{p}_3)$ (d).}
\end{figure}

The polyspectra of the displacement field in LPT can be evaluated in
almost the same way as in the case of density fields in Eulerian
perturbation theory. Substituting Eqs.~(\ref{eq:1-11}),
(\ref{eq:1-19}), and (\ref{eq:1-21}) into Eq.~(\ref{eq:1-10}), the
perturbative expansion of the polyspectra is calculated. We assume the
initial density field is a random Gaussian field, in which case the
cumulants of the linear density field are completely expressible by a
linear power spectrum, $P_{\rm L}(k) = D^2 P_0(k)$. Up to second order
in $P_{\rm L}(k)$, we obtain
\begin{align}
&  C_{ij}(\bm{p}) = C^{(11)}_{ij}(\bm{p})
  + C^{(22)}_{ij}(\bm{p}) + C^{(13)}_{ij}(\bm{p})
  + C^{(31)}_{ij}(\bm{p})
\nonumber\\
&\qquad\qquad + \cdots, 
\label{eq:1-24a}\\
&  C_{ijk}(\bm{p}_1,\bm{p}_2,\bm{p}_3)
   = C^{(112)}_{ijk}(\bm{p}_1,\bm{p}_2,\bm{p}_3)
\nonumber\\
&\qquad
   + C^{(121)}_{ijk}(\bm{p}_1,\bm{p}_2,\bm{p}_3)
   + C^{(211)}_{ijk}(\bm{p}_1,\bm{p}_2,\bm{p}_3)
   + \cdots,
\label{eq:1-24b}
\end{align}
where we define quantities $C^{(nm)}_{ij}$, $C^{(nml)}_{ijk}$ by
\begin{align}
&  \left\langle \tilde{\Psi}^{(n)}_i(\bm{p}) \tilde{\Psi}^{(m)}_j(\bm{p}')
   \right\rangle_{\rm c} = (2\pi)^3 \delta^3(\bm{p} + \bm{p}')\,
   C^{(nm)}_{ij}(\bm{p}),
\label{eq:1-24-1a}\\
&  \left\langle
    \tilde{\Psi}^{(n)}_i(\bm{p}_1) \tilde{\Psi}^{(m)}_j(\bm{p}_2)
    \tilde{\Psi}^{(l)}_k(\bm{p}_3)
   \right\rangle_{\rm c}
\nonumber\\
& \qquad   = (2\pi)^3 \delta^3(\bm{p}_1 + \bm{p}_2 + \bm{p}_3)\,
   C^{(nml)}_{ijk}(\bm{p}_1,\bm{p}_2,\bm{p}_3),
\label{eq:1-24-1b}
\end{align}
which are explicitly given by using LPT kernels as
\begin{align}
&  C^{(11)}_{ij}(\bm{p})
   = L^{(1)}_i(\bm{p}) L^{(1)}_j(\bm{p}) P_{\rm L}(p),
\label{eq:1-25a}\\
&  C^{(22)}_{ij}(\bm{p})
   = \frac12 \int \frac{d^3p'}{(2\pi)^3}
   L^{(2)}_i(\bm{p}',\bm{p}-\bm{p}')
   L^{(2)}_j(\bm{p}',\bm{p}-\bm{p}')
\nonumber\\
&\qquad\qquad\qquad\qquad\qquad\qquad \times
   P_{\rm L}(p') P_{\rm L}(|\bm{p} - \bm{p}'|),
\label{eq:1-25b}\\
&  C^{(13)}_{ij}(\bm{p}) = C^{(31)}_{ji}(\bm{p})
\nonumber\\
&\,\,
   = \frac12 L^{(1)}_i(\bm{p}) P_{\rm L}(p)
   \int \frac{d^3p'}{(2\pi)^3}
   L^{(3)}_j(\bm{p},-\bm{p}',\bm{p}') P_{\rm L}(p'),
\label{eq:1-25c}
\end{align}
and
\begin{align}
&  C^{(112)}_{ijk}(\bm{p}_1,\bm{p}_2,\bm{p}_3)
   = C^{(211)}_{kij}(\bm{p}_3,\bm{p}_1,\bm{p}_2)
\nonumber\\
&\quad
   = C^{(121)}_{jki}(\bm{p}_2,\bm{p}_3,\bm{p}_1)
\nonumber\\
&\quad
   = - L^{(1)}_i(\bm{p}_1) L^{(1)}_j(\bm{p}_2)
   L^{(2)}_k(\bm{p}_1,\bm{p}_2) P_{\rm L}(p_1) P_{\rm L}(p_2).
\label{eq:1-26}
\end{align}
The polyspectra of order four or higher do not contribute up to second
order in $P_{\rm L}(k)$. In random Gaussian fields, it can generally
be shown that $N$-polyspectrum has the order $N-1$ in $P_{\rm L}(k)$,
for the connectedness of the cumulants. In SPT, diagrammatic
techniques prove to be quite useful \cite{Goroff86,SF96}. Similarly,
we find diagrammatic representations of the above quantities of
Eqs.~(\ref{eq:1-25a})--(\ref{eq:1-26}), which are given in
Fig.~\ref{fig:polydiag}.

\begin{figure}
\includegraphics[width=20pc]{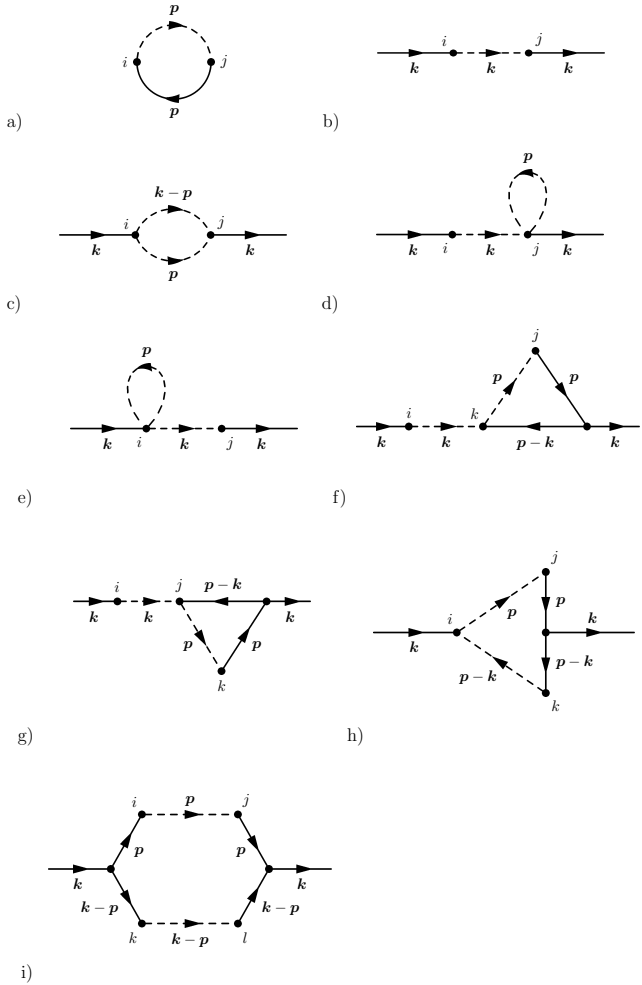}
\caption{\label{fig:psdiag} Diagrams for our resummed power spectrum
  of Eq.~(\ref{eq:1-27}).}
\end{figure}

\begin{figure}
\vspace{2em}
\includegraphics[width=15pc]{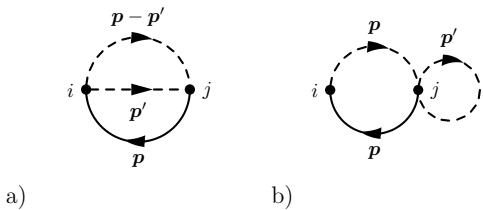}
\caption{\label{fig:twobubble} Two-loop bubble diagrams (see text).}
\end{figure}

At this point, Eq.~(\ref{eq:1-8}) is expanded by LPT by using
Eqs.~(\ref{eq:1-14a}), (\ref{eq:1-14b}), and
(\ref{eq:1-25a})--(\ref{eq:1-26}). As promised, only quantities
$B^{(N)}$ are expanded from the exponent in Eq.~(\ref{eq:1-8}) and
keep $A^{(N)}$ in the exponent. The result is
\begin{align}
&  P(\bm{k}) =
   \exp\left[
       - k_i k_j \int \frac{d^3p}{(2\pi)^3} C^{(11)}_{ij}(\bm{p})
    \right]
\nonumber\\
& \times
  \Biggl\{
    k_i k_j \left[
        C^{(11)}_{ij}(\bm{k}) + C^{(22)}_{ij}(\bm{k})
        + C^{(13)}_{ij}(\bm{k}) + C^{(31)}_{ij}(\bm{k})
    \right]
\nonumber\\
& \qquad
    + k_i k_j k_k
      \int \frac{d^3p}{(2\pi)^3}
      \left[
         C^{(112)}_{ijk}(\bm{k},-\bm{p},\bm{p}-\bm{k})
     \right.
\nonumber\\
& \qquad\quad
     \left. +\,
         C^{(121)}_{ijk}(\bm{k},-\bm{p},\bm{p}-\bm{k})
         + C^{(211)}_{ijk}(\bm{k},-\bm{p},\bm{p}-\bm{k})
      \right]
\nonumber\\
& \qquad
    + \frac12 k_i k_j k_k k_l
      \int \frac{d^3p}{(2\pi)^3}
      C^{(11)}_{ij}(\bm{p}) C^{(11)}_{kl}(\bm{k}-\bm{p})
  \Biggr\}.
\label{eq:1-27}
\end{align}
Except for the linear-theory contribution, $C^{(11)}_{ij}(\bm{k})$,
all the polyspectra in the above terms are accompanied by integrations
over wave vectors. Accordingly, these terms correspond to the 1-loop
diagrams in Fig.~\ref{fig:psdiag}.
The bubble diagram (a) of Fig.~\ref{fig:psdiag} corresponds to the
exponent of the first factor. The tree diagram (b) corresponds to the
linear contribution $C^{(11)}_{ij}$, and loop diagrams (c)--(e)
correspond to $C^{(22)}_{ij}$, $C^{(13)}_{ij}$, and $C^{(31)}_{ij}$,
respectively. The loop diagrams (f)--(h) correspond to integrals of
$C^{(112)}_{ijk}$, $C^{(121)}_{ijk}$, and $C^{(211)}_{ijk}$,
respectively. The last loop diagram i) corresponds to the last
integral in Eq.~(\ref{eq:1-27}). The expansion of Eq.~(\ref{eq:1-27})
is therefore a loop expansion, and all the 1-loop diagrams are
exhausted in this equation. In the exponent of the prefactor, we could
consider the higher-order terms such as integrals of $C^{(22)}_{ij}$,
$C^{(13)}_{ij}$. These terms correspond to the 2-loop diagrams of
Fig.~\ref{fig:twobubble}.
Although those terms are second order in $P_{\rm L}(k)$, their
contributions to the total power spectrum are third order since they
are multiplied by first- or higher-order terms in Eq.~(\ref{eq:1-27}).
Therefore, we do not include these 2-loop contributions and truncate
all the multiloop contributions.

\section{\label{sec:RealSpace}
Clustering in Real Space
}

\subsection{\label{subsec:PSReal}
The power spectrum
}

It is straightforward to calculate Eq.~(\ref{eq:1-27}) using the
expression of the polyspectra of the displacement field,
Eqs.~(\ref{eq:1-25a})--(\ref{eq:1-26}), and perturbation kernels,
Eqs.~(\ref{eq:1-22a})--(\ref{eq:1-23}). Contractions with spatial
indices in Eq.~(\ref{eq:1-27}) can be taken before evaluating the
momentum integrations. The transverse part in the third-order kernel,
Eq.~(\ref{eq:1-22c}), vanishes in course of calculation, because of
the inner product with vector $\bm{k}$. As a result, the calculation
is quite similar to that in SPT, where some of the angular
integrations can be analytically done \cite{MSS}. The result is
\begin{widetext}
\begin{align}
  P(k) =
&
  \exp\left[
      -\frac{k^2}{6\pi^2} \int dp\, P_{\rm L}(p)
  \right]
\nonumber\\
& \times
  \Biggl\{
    P_{\rm L}(k)
    + \frac{1}{98} \frac{k^3}{4\pi^2}
      \int_0^\infty dr P_{\rm L}(kr) \int_{-1}^1 dx
      P_{\rm L}[k\,(1 + r^2 - 2rx)^{1/2}]
      \frac{(3r + 7x - 10rx^2)^2}{(1 + r^2 - 2rx)^2}
\nonumber\\
& \qquad\qquad  + \frac{1}{252} \frac{k^3}{4\pi^2} P_{\rm L}(k) 
      \int_0^\infty dr P_{\rm L}(kr) 
      \left[\frac{12}{r^2} + 10 + 100 r^2 - 42 r^4
          + \frac{3}{r^3}(r^2 - 1)^3 (7r^2 + 2)
            \ln\left|\frac{1+r}{1-r}\right| \right]
  \Biggr\}.
\label{eq:2-1}
\end{align}
\end{widetext}
Comparing this result with that of 1-loop SPT \cite{MSS}, one
immediately notices the similarity of the equations. The only
differences are the existence of the exponential prefactor and the
number $+10$ of the second term in the last square bracket, which is
$-158$ in SPT instead. Thus, Eq.~(\ref{eq:2-1}) is equivalently
represented as
\begin{align}
&  P(k) =
  \exp\left[
      -\frac{k^2}{6\pi^2} \int dp\, P_{\rm L}(p)
  \right]
\nonumber\\
& \quad \times
  \left[ P_{\rm L}(k) + P_{\rm SPT}^{\mbox{\scriptsize 1-loop}}(k)
      +
      \frac{k^2}{6\pi^2} P_{\rm L}(k) \int dp\, P_{\rm L}(p)
  \right],
\label{eq:2-1-1}
\end{align}
where $P_{\rm SPT}^{\mbox{\scriptsize 1-loop}}(k)$ is a pure 1-loop
contribution (without a linear contribution) to the power spectrum in
SPT. It is obvious from this expression that expanding the exponential
prefactor exactly reproduces the SPT result at the 1-loop level. The
exponential prefactor contains information from an infinite series of
higher-order perturbations in terms of SPT. In terms of diagrams of
SPT \cite{Goroff86}, the exponential prefactor comes from the diagrams
in which two points are connected by a single internal line with
$P_{\rm L}(k)$ but vertices are dressed by loops. However, those
diagrams are not completely resummed, and parts of the resulting terms
are resummed. Therefore, the resummation of this paper corresponds to
a partial vertex renormalization in terms of the SPT.

On the other hand, the resulting expression has a certain similarity
with RPT \cite{CS06a,CS06b}, because of the exponential prefactor. In
fact, a square root of this factor coincides with the nonlinear
propagator in a large-$k$ limit of RPT. In its formulation, the
nonlinear power spectrum is divided into two parts. The first part is
linearly dependent on the initial power spectrum with a coefficient
given by the square of the propagator. The second part corresponds to
contributions from mode couplings among different scales. RPT uses
loop expansions to evaluate the mode-coupling contributions. Our
expression has a similar structure, although the methods are quite
different. One possible reason for this similarity is that the RPT
propagator in the Zel'dovich approximation exactly coincides with the
square root of the exponential prefactor of Eq.~(\ref{eq:2-1}) above.
In the Zel'dovich approximation, only $C^{(11)}_{ij}$ is present for
cumulants of the displacement field. As a result, loop corrections
come only from diagrams of type i) in Fig.~\ref{fig:psdiag} and its
multiloop extensions ($n$-loop diagrams with $n+1$ internal lines
connected to the two external vertices), in which the vertex
renormalization cannot appear. As a result, RPT and our approach turn
out to be equivalent in the Zel'dovich approximation (see the
introduction of Ref.~\cite{CS06a}).

Having an exponential damping factor, the power spectrum in our
approximations does not have power at large wave numbers. This
behavior is not physical in the fully nonlinear regime, since the
nonlinear evolution enhances the power spectrum on small scales. Our
approximations only apply to the quasilinear regime on large scales.
An advantage of this property is that the integration of the power
spectrum is stable so that other statistical quantities may be
obtained, such as the correlation function, on quasilinear scales.

On the other hand, our approximations for the power spectrum itself
fail when the exponential damping factor is efficient. It is useful to
have an estimate for the validity range of wave numbers in our
approximations. Since the damping scale is given by
\begin{equation}
  k_{\rm NL} \equiv 
  \left[\frac{1}{6\pi^2}\int dp\, P_{\rm L}(p) \right]^{-1/2},
\label{eq:2-1-2}
\end{equation}
a possible criterion for the validity range is $k < \alpha k_{\rm
  NL}$, where $\alpha$ should be determined according to the required
accuracy. We will see below that $\alpha = 0.5$ corresponds to the
accuracy within a few percent, compared to the $N$-body simulations.
Throughout the rest of this paper, we consider $k < k_{\rm NL}/2$ as
the validity range in our approximations.

\begin{figure}
\includegraphics[width=20pc]{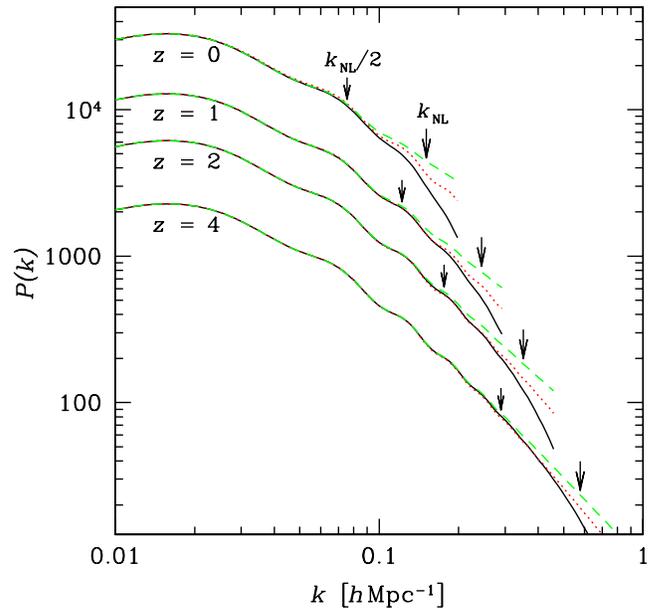}
\caption{\label{fig:logps} Comparison of power spectra by different
  approximations at redshifts $z=0, 1, 2, 4$ (from upper to lower
  lines). {\em Black (solid) line}: this work; {\em red (dotted)
    line}: linear theory; {\em green (dashed) line}: 1-loop SPT. Two
  nonlinear scales $k_{\rm NL}$, $k_{\rm NL}/2$ in each redshift are
  indicated by arrows. }
\end{figure}

In Fig.~\ref{fig:logps}, the power spectra of our result, and those of
linear theory and 1-loop SPT, are compared.
The linear power spectrum $P_{\rm L}(k)$ is calculated from the output
of CMBFAST \cite{CMBFAST} at $z=0$. We adopt a cosmological model with
parameters $\Omega_{\rm m} = 0.27$, $\Omega_{\rm \Lambda} = 0.73$,
$\Omega_{\rm b} = 0.046$, $h = 0.72$, $n_{\rm s} = 1$ and $\sigma_8 =
0.9$. The damping scale $k_{\rm NL}$ and the claimed validity scale
$k_{\rm NL}/2$ are indicated in the figure. It is clearly seen that
our power spectra are expnentially damped on small scales $k > k_{\rm
  NL}$. In this log-log plot, differences among the three
approximations in our claimed validity range $k < k_{\rm NL}/2$ are
hardly seen. Our approach is advantageous when a very accurate
estimation on large scales is crucial. The determination of the BAO
scale really requires such accuracy.

\begin{figure*}
\begin{center}
\includegraphics[width=32pc]{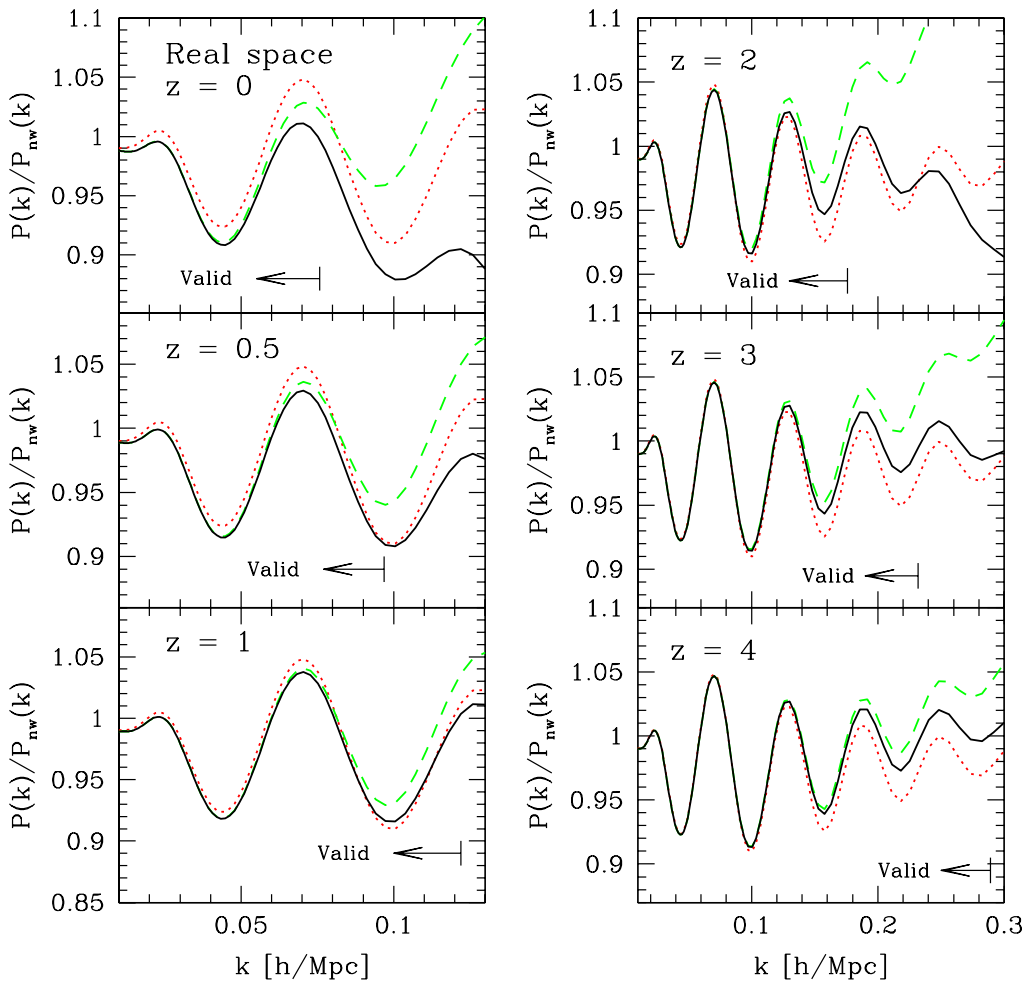}
\caption{\label{fig:realpsbao} Nonlinear evolution of the baryon
  acoustic oscillations in real space for various redshifts, $z = 0$
  (top left), $0.5$ (center left), $1$ (bottom left), $2$ (top right),
  $3$ (center right), $4$ (bottom right). Each power spectrum is
  divided by a smoothed, no-wiggle linear power spectrum $P_{\rm
    nw}(k)$ \cite{EH99}. {\em Black (solid) line}: this work; {\em red
    (dotted) line}: linear theory; {\em green (dashed) line}: 1-loop
  SPT. The validity range $k < k_{\rm NL}/2$, where our result is
  expected to be accurate within a few percent, is indicated by
  arrows.}
\end{center}
\end{figure*}

In Fig.~\ref{fig:realpsbao}, detailed features in nonlinear power
spectra are shown on BAO scales.
To enhance the baryon oscillations, each power spectrum is divided by
a smoothed, no-wiggle linear power spectrum $P_{\rm nw}(k)$
\cite{EH99}. The same cosmological parameters are assumed as in
Fig.~\ref{fig:logps}. The validity ranges are indicated in the figure.
There are differences among three approximations in our validity
ranges.

\begin{figure}
\begin{center}
\includegraphics[width=20pc]{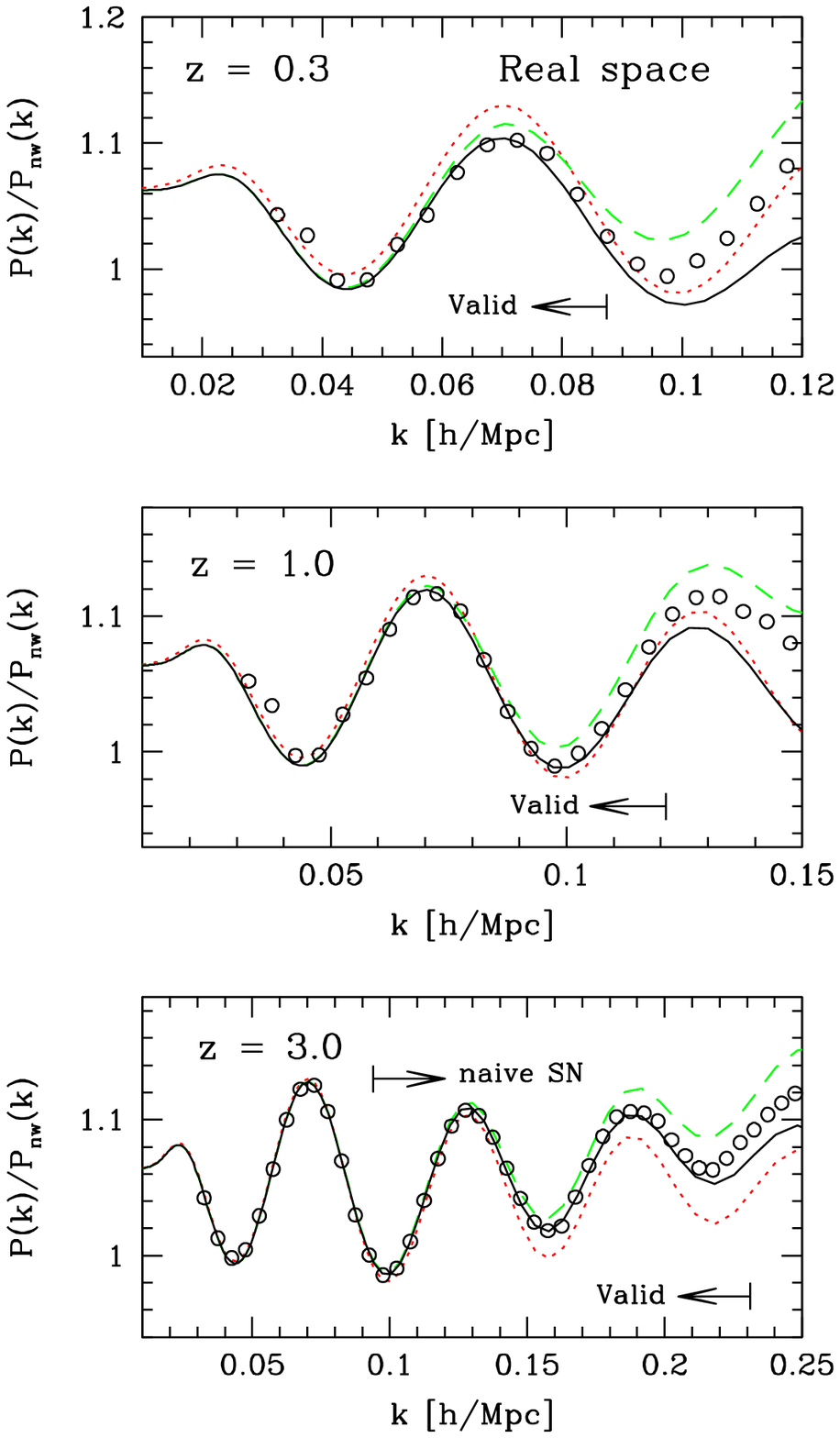}
\caption{\label{fig:psSE} Comparison of the power spectra to the
  $N$-body simulations of Ref.~\cite{SE05} in real space at redshifts
  $z=0.3$, $1$, and $3$ from top to bottom, respectively. {\em Open
    circles}: $N$-body simulations; {\em Black (solid) line}: this
  work; {\em red (dotted) line}: linear theory; {\em green (dashed)
    line}: 1-loop SPT. Only nonlinear deviations from the linear
  growth is measured in $N$-body simulations to reduce cosmic
  variances. Estimates of validity range, $k < k_{\rm NL}/2$ are
  indicated by arrows. In the $z=3$ sample, the range where
  contributions from the na\"ive shot noise exceed $1\%$ is also
  indicated, althogh the true shot noise could be smaller because of
  the regularity of particle positions at such a large redshift in the
  simulation.}
\end{center}
\end{figure}

In Fig.~\ref{fig:psSE}, our result of the power spectrum in real space
is compared with the $N$-body simulations of Ref.~\cite{SE05} using
the same parameters as Seo and Eisenstein's.
In this case, $n_{\rm s}=0.99$ and other cosmological parameters are
the same as above. The linear power spectrum is calculated from the
output of CMBFAST at $z=49$ to match the input power spectrum of the
simulations. Since the no-wiggle power spectrum $P_{\rm nw}(k)$
matches to the power spectrum at $z=0$, the baseline is slightly
larger than unity, because of nongravitational effects in the transfer
function. To reduce the effects of cosmic variance in $N$-body
simulations, we calculate the differences (in logarithmic scales)
between the output power spectra and linearly evolved initial power
spectra at $z=49$, both measured in simulations. The resulting
differences are added to the linear theory prediction in the figure.
In this way, initial deviations of the power spectrum from the
theoretical power spectrum are subtracted (similar presentations are
seen in Ref.~\cite{Spr05}). The actual output power spectrum of
simulations is noisier than that in the figure especially on larger
scales. As described in Ref.~\cite{SE05}, the Savitzky-Golay filtering
is applied when the power spectrum in each $N$-body sample is
evaluated. The shot noise is not corrected for the $N$-body samples,
since the initial positions of particles are on regular grids, and the
na\"ive shot noise $P_{\rm SN}(k) = 1/\bar{n}$, where $\bar{n}$ is the
number density of particles, does not apply at a large redshift in the
simulation \cite{Seo07}. It is not obvious how to correct for the shot
noise in such a situation. The samples with $z=0.3$ and $z=1$ are
hardly affected by the shot noise because the signal amplitudes are
large, but contributions from the na\"ive shot noise in the $z=3$
sample are over $1\%$ for $k > 0.094 h\,{\rm Mpc}^{-1}$. Therefore,
the $z=3$ sample could be more or less affected by nontrivial shot
noise in that range, although that is smaller than the na\"ive shot
noise.

In all cases of the three redshifts $z=0.3$, $1$, $3$, the ranges
where our results and simulations agree are broader than those of
1-loop SPT. Overall, the 1-loop SPT overestimates the power spectrum
on relatively small scales, while our result is always better on the
same scales. For example, the height of the peak at $k \simeq
0.07\,h\,{\rm Mpc}^{-1}$ of the $z=0.3$ sample is explained by our
work, while the 1-loop SPT overestimates the height. In a range $k
\simeq 0.09$--$0.13\, h\,{\rm Mpc}^{-1}$ of the $z=1$ sample, $N$-body
data are (accidentally) closer to the linear theory rather than the
1-loop SPT. This tendency is also explained by our work. In the $z=3$
sample, our result matches the $N$-body result in the range $k <0.2\,
h\,{\rm Mpc}^{-1}$, while linear theory and 1-loop SPT do not (under
the condition that the nontrivial shot noise is negligible as noticed
above). The simulations and our results agree within a few percent in
our claimed validity range, $k < k_{\rm NL}/2$.

\subsection{\label{subsec:CFReal}
The correlation function}

The correlation function $\xi(r)$ is given by a Fourier transform of
the power spectrum. After angular integrations, they are related by
\begin{equation}
    \xi(r) =
    \int_0^\infty \frac{k^2 dk}{2\pi^2} j_0(kr) P(k)
\label{eq:2-2}
\end{equation}
where $j_0(x) = x^{-1}\sin x$ is the spherical Bessel function of
zeroth order.

In the 1-loop SPT, the power spectrum badly behaves on small scales to
perform this integral. The 1-loop contribution to the power spectrum
in SPT consists of two terms, $P_{\rm SPT}^{\mbox{\scriptsize
    1-loop}}(k) = P_{22}(k) + P_{13}(k)$. Each term asymptotically
behaves as \cite{vish83, MSS}
\begin{equation}
    P_{22}(k) \simeq - P_{13}(k) \simeq
    \frac{ k^2 P_{\rm L}(k)}{6 \pi^2}
    \int_0^\infty dp P_{\rm L}(p),
    \ \ (k \rightarrow \infty),
\label{eq:2-2-1}
\end{equation}
and the net power spectrum is much smaller in magnitude than each
term. On mildly small scales, the residuals also behave as $\propto
k^2 P_{\rm L}(k)$ for a CDM-like power spectrum. Therefore, the power
spectrum on small scales contributes to the integral in
Eq.~(\ref{eq:2-2}) even if the separation $r$ is large. In a
small-scale limit, $k \rightarrow \infty$, the leading contributions
of the two terms are exactly canceled out, and the net contribution
behaves as \cite{vish83, MSS}
\begin{equation}
    P_{\rm SPT}^{\mbox{\scriptsize 1-loop}}(k) \propto
    P_{\rm L}(k) \int_0^\infty dp p^2 P_{\rm L}(p),
    \quad (k \rightarrow \infty).
\label{eq:2-2-2}
\end{equation}
For a CDM-like power spectrum, $d\ln P_{\rm L}(k)/d\ln k > -3$, and
the integral in Eq.~(\ref{eq:2-2-2}) diverges. As a result, one cannot
evaluate the correlation function by the 1-loop SPT, because the
integral of Eq.~(\ref{eq:2-2}) does not converge. This divergence
obviously reflects the fact that the 1-loop SPT is invalid on small
scales. Since the correlation function is an observable quantity, this
is an undesirable property in SPT. Such a problem does not occur in
our expression of Eq.~(\ref{eq:2-1}): the power spectrum is
exponentially suppressed in the small-scale limit. This behavior is
not physical on sufficiently small scales, since our expression is
applicable only to the weakly nonlinear regime. However, small-scale
behavior in the power spectrum should not be relevant to the
correlation function on large scales, since dominant contributions of
the power spectrum to the correlation function of Eq.~(\ref{eq:2-2})
come from scales $k \sim r^{-1}$ for a CDM-like power spectrum.

Since the exponential prefactor in Eq.~(\ref{eq:2-1}) acts as a
Gaussian damping function, the resulting correlation function has a
form of convolution by a Gaussian smoothing function. This smoothing
effect by nonlinearity is already known, which smears the BAO peak in
the correlation function \cite{ESW07}. The physical origin of this
nonlinear smearing is the random motion of mass elements, because the
dominant contribution of this smearing factor comes from $\langle
|\bm{\Psi}|^2 \rangle$ as obviously seen by Eqs.~(\ref{eq:1-8}) and
(\ref{eq:1-9a}). It is worth noting that this nonlinear smearing
effect is not captured by linear theory and higher-order SPT.

\begin{figure*}
\begin{center}
\includegraphics[width=32pc]{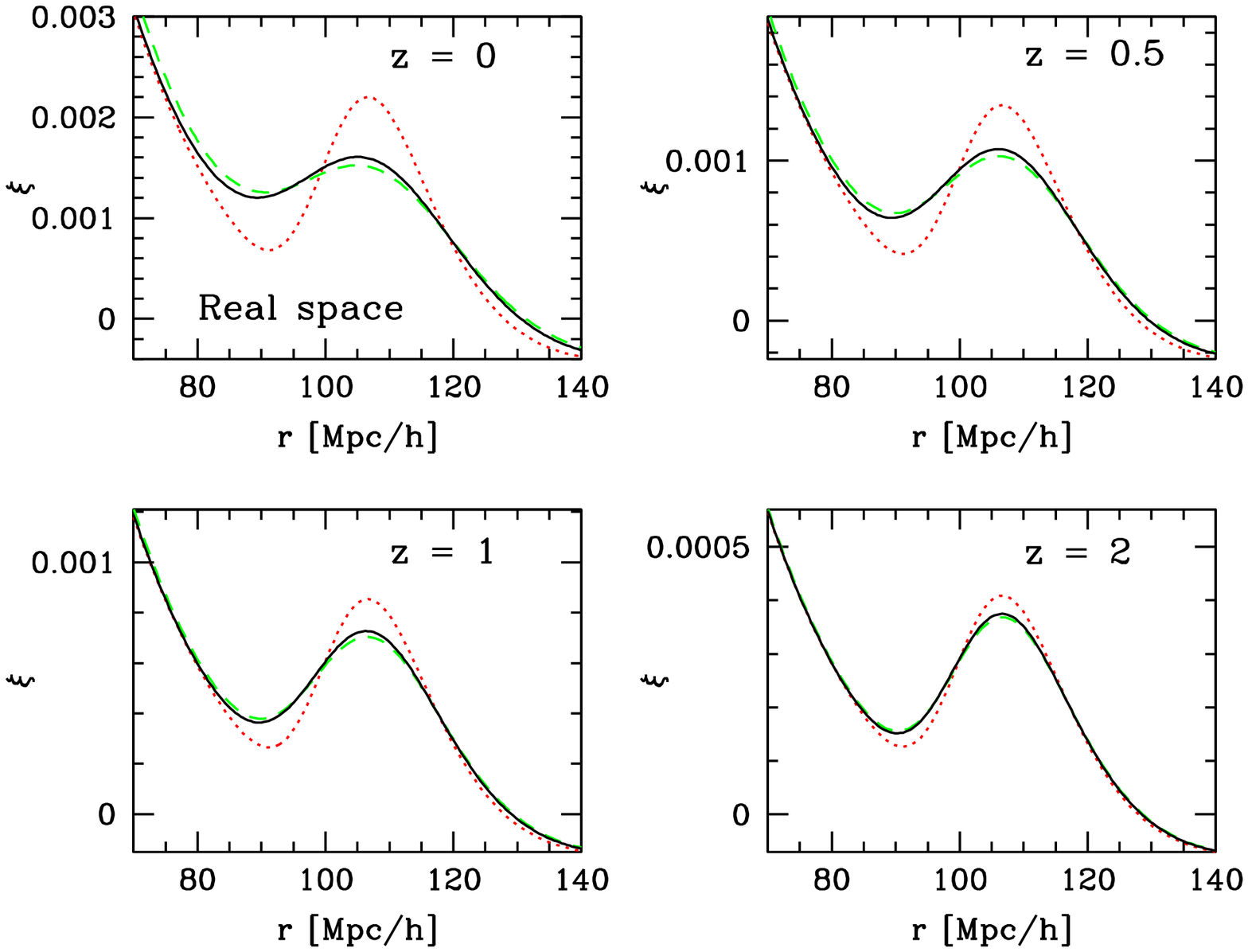}
\caption{\label{fig:realxibao} Nonlinear evolution of the baryon
  acoustic peak in real-space correlation function for various
  redshifts, $z = 0$ (top left), $0.5$ (top right), $1$ (bottom left),
  $2$ (bottom right). {\em Black (solid) line}: this work; {\em red
    (dotted) line}: linear theory; {\em green (dashed) line}:
  Gaussian-filtered linear theory. The 1-loop SPT cannot predict
  the correlation function.}
\end{center}
\end{figure*}

In Fig.~\ref{fig:realxibao}, the nonlinear correlation function is
plotted on BAO scales.
The same cosmological parameters as in Fig.~\ref{fig:realpsbao} are
adopted. Nonlinear smearing effects on the baryon acoustic peak are
described by our theory. In Fig.~\ref{fig:realxibao},
Gaussian-filtered linear theory is overplotted. The filtering length
is given by the one that the exponential prefactor in
Eq.~(\ref{eq:2-1}) implies:
\begin{equation}
  R_{\rm smear} = 
  \left[\frac{1}{6\pi^2}\int dp\, P_{\rm L}(p) \right]^{1/2}.
\label{eq:2-3}
\end{equation}
This simple model approximately explains the nonlinear smearing
effects. In detail, this simple model overestimates the small-scale
clustering and height of the trough, and underestimates the peak
height.

\begin{figure}
\begin{center}
\includegraphics[width=18pc]{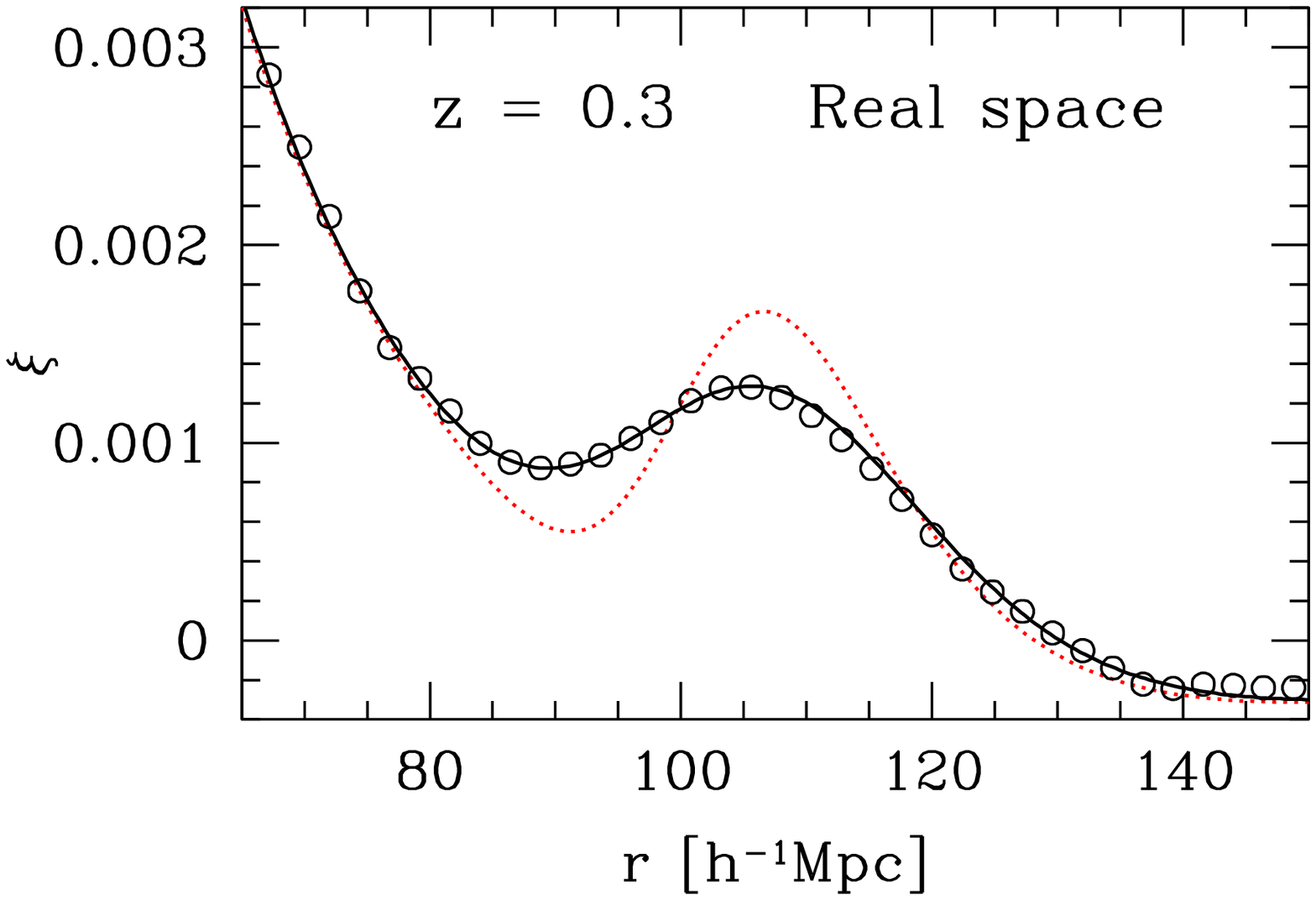}
\caption{\label{fig:xiSE} Comparison of the correlation functions to
  the $N$-body simulations of Refs.~\cite{SE05,ESSS07} in real space.
  {\em Open circles}: $N$-body results; {\em Black (solid) line}: this
  work; {\em red (dotted) line}: linear theory. Only nonlinear
  deviations from the linear growth are measured in $N$-body
  simulations to reduce finiteness effects. }
\end{center}
\end{figure}

In Fig.~\ref{fig:xiSE}, our result of the correlation function in real
space at $z=0.3$ is compared with the $N$-body simulations of
Refs.~\cite{SE05,ESSS07} using the same parameters as used in these
references.
As in the case of the power spectrum, we compute differences (in
linear scales) between the output correlation function and the
linearly evolved initial correlation function at $z=49$ in the
simulation, and the resulting differences are added to the
linear-theory prediction in the figure. If we directly plot the
measured correlation function, the amplitudes on $r \simeq
70$--$100\,h^{-1}{\rm Mpc}$ and $115$--$135\,h^{-1}{\rm Mpc}$ are
slightly (about $10^{-4}$) larger. This tendency is already seen even
in the initial correlation function at $z=49$, which can be identified
to the finiteness effect of the simulations. Therefore, we only
measure the nonlinear deviations from the linear theory in the
simulation.

The agreement between the simulation and our result is quite good.
There is not any fitting parameter at all. In the corresponding power
spectrum, the $z=0.3$ plot in Fig.~\ref{fig:psSE}, the difference
between our result and that of 1-loop SPT may seem to be small. In the
correlation function, however, the difference is really big, since the
1-loop SPT cannot predict the correlation function, while our result
explains the nonlinear effects on BAO scales very precisely.

\section{\label{sec:RedshiftSpace}
Clustering in Redshift Space}

\subsection{\label{subsec:RealtoRed}
Mapping from real space to redshift space}

Our approach above can be extensible to the calculation in redshift
space. In redshift surveys, the position of each galaxy is measured by
the redshift. In a completely homogeneous universe, the comoving
distance--redshift relation is given by a standard formula for the
Friedmann-Lema\^itre universe:
\begin{equation}
  x(z) = \int_0^z \frac{cz}{H(z)},
\label{eq:3-1}
\end{equation}
where $H(z)$ is the time-dependent Hubble parameter. In reality,
however, the Universe is locally inhomogeneous, and the comoving
distance to a galaxy of redshift $z$ is not exactly given by
Eq.~(\ref{eq:3-1}). For the galaxy redshifts, the leading source of
the deviation is the Doppler shift caused by the peculiar velocities
of the galaxy. The observed redshift $z_{\rm obs}$ with the effect of
the peculiar velocity can be calculated by solving the geodesic
equation. The comoving distance in redshift space is defined by $s =
x(z_{\rm obs})$, where, in the RHS, the redshift-distance relation of
Eq.~(\ref{eq:3-1}) is adopted even though we live in an inhomogeneous
universe. When the peculiar velocities are nonrelativistic, the
comoving distance in real space, $x$, and that in redshift space $s$
is related by $s = x + v_z/aH$ \cite{TM2000}, where $v_z =
\bm{v}\cdot\hat{\bm{z}}$ is a line-of-sight component of the peculiar
velocity $\bm{v}$. Throughout this paper, we work in
``distant-observer'' or ``plane-parallel'' approximation, where the
line-of-sight direction is fixed in space and denoted by
$\hat{\bm{z}}$. We ignore the contribution from the peculiar velocity
of the observer, which does not contribute to the following
calculation of the power spectrum. When the curvature scale of the
Universe is much larger than clustering scales we are interested in,
we can safely use local Cartesian coordinates in the volume of galaxy
surveys. Thus the relation between the position in real space $\bm{x}$
and that in redshift space $\bm{s}$ is given by
\begin{equation}
  \bm{s} = \bm{x}
  + \frac{\hat{\bm{z}}\cdot\bm{v}}{aH}\, \hat{\bm{z}},
\label{eq:3-2}
\end{equation}
where $\bm{v} = a\dot{\bm{x}}$. Therefore, using Eq.~(\ref{eq:1-1}),
the displacement field in redshift space is given by
\begin{equation}
  \bm{\Psi}^{\rm s} = 
  \bm{\Psi} + \frac{\hat{\bm{z}}\cdot\dot{\bm{\Psi}}}{H}\, \hat{\bm{z}}.
\label{eq:3-3}
\end{equation}
A similar equation has been applied to the analysis of the Zel'dovich
approximation \cite{TH96}.

In the time-independent approximation to the perturbative kernels,
$\bm{\Psi}^{(n)} \propto D^n$, and therefore, the time derivative of
the displacement field is simply
\begin{equation}
  \dot{\bm{\Psi}}^{(n)} = n H f \bm{\Psi}^{(n)},
\label{eq:3-4}
\end{equation}
where $f = d\ln D/d\ln a =(HD)^{-1}\dot{D}$ is the logarithmic
derivative of the linear growth rate. Thus, the displacement field of
each perturbation order in redshift space is given by
\begin{equation}
 \bm{\Psi}^{{\rm s}(n)} = 
  \bm{\Psi}^{(n)}
  + n f (\hat{\bm{z}}\cdot\bm{\Psi}^{(n)})\, \hat{\bm{z}},
\label{eq:3-5}
\end{equation}
which is just a linear mapping of the displacement vector of each
order. This linear transformation is characterized by a redshift-space
distortion tensor $R^{(n)}_{ij}$ for $n$th order perturbations,
defined by
\begin{equation}
  R^{(n)}_{ij} = \delta_{ij} + n f \hat{z}_i \hat{z}_j,
\label{eq:3-6}
\end{equation}
with which Eq.~(\ref{eq:3-5}) reduces to $\Psi^{{\rm s}(n)}_i =
R^{(n)}_{ij}\Psi^{(n)}_j$, or in a vector notation, $\bm{\Psi}^{{\rm
    s}(n)} = R^{(n)}\bm{\Psi}^{(n)}$. Therefore, the perturbative
kernels in redshift space are given simply by
\begin{equation}
  \bm{L}^{{\rm s}(n)} = R^{(n)} \bm{L}^{(n)}.
\label{eq:3-7}
\end{equation}
Thus, our calculation in real space can be simply generalized to that
in redshift space by using the redshift-space perturbative kernels in
Eqs.~(\ref{eq:1-25a})--(\ref{eq:1-26}), while the form of
Eq.~(\ref{eq:1-27}) is unchanged in redshift space, provided that the
cumulants of the displacement field are evaluated in redshift space.
The mappings of the order-by-order cumulants are quite simple:
\begin{align}
&  C^{{\rm s}(nm)}_{ij}
   = R^{(n)}_{ik} R^{(m)}_{jl} C^{(nm)}_{kl}
\label{eq:3-8a}\\
&  C^{{\rm s}(n_1 n_2 n_3)}_{i_1 i_2 i_3}
   = R^{(n_1)}_{i_1 j_1} R^{(n_2)}_{i_2 j_2} R^{(n_3)}_{i_3 j_3}
   C^{(n_1 n_2 n_3)}_{j_1 j_2 j_3}.
\label{eq:3-8b}
\end{align}

\subsection{\label{subsec:PSRed}
The power spectrum in redshift space}

The calculation of Eq.~(\ref{eq:1-27}) in redshift space is more
complicated than in real space, since the anisotropy is introduced by
the redshift-space distortion tensors. Accordingly, the momentum
integrations should be evaluated before taking inner products with the
vector $\bm{k}$. In redshift space, however, the cumulants
$C^{(nm)}_{ij}$ and $C^{(n_1 n_2 n_3)}_{ijk}$ in Eq.~(\ref{eq:1-27})
should be replaced by redshift-space counterparts, $C^{{\rm
    s}(nm)}_{ij}$ and $C^{{\rm s}(n_1 n_2 n_3)}_{ijk}$ of
Eqs.~(\ref{eq:3-8a}) and (\ref{eq:3-8b}). Since the redshift
distortion tensors $R^{(n)}_{ij}$ are anisotropic, the integrands do
not reduce to scalar functions. Therefore we need to explicitly
evaluate integrals of tensors. The results are given in
Appendix~\ref{app:IntTens}, Eqs.~(\ref{eq:a-4a})--(\ref{eq:a-4c}) and
(\ref{eq:a-6a})--(\ref{eq:a-6c}). Using those equations, it is
straightforward to obtain an expression of the power spectrum $P_{\rm
  s}(\bm{k})$ in redshift space. There are many terms in the result,
which are arranged as
\begin{align}
&  P_{\rm s}(\bm{k}) =
  \exp\left\{ - k^2[1 + f(f+2)\mu^2] A \right\}
\nonumber\\
& \qquad\qquad \times
  \left[ (1 + f\mu^2)^2 P_{\rm L}(k)
      + \sum_{n,m} \mu^{2n} f^m E_{nm}(k)
  \right],
\label{eq:3-9}
\end{align}
where $\mu = \hat{\bm{z}}\cdot \bm{k}/k$ is the direction cosine of
the wave vector $\bm{k}$ with respect to the line of sight,
\begin{equation}
  A = \frac{1}{6\pi^2} \int dp P_{\rm L}(p),
\label{eq:3-10}
\end{equation}
and
\begin{align}
&  E_{00} = \frac{9}{98} Q_1 + \frac37 Q_2 + \frac12 Q_3 +
\frac{10}{21} R_1 + \frac67 R_2,
\label{eq:3-11a}\\
&  E_{11} = 4 E_{00},
\label{eq:3-11b}\\
&  E_{12} = -\frac{3}{14} Q_1 - \frac32 Q_2 + \frac14 Q_4 - \frac67 R_1,
\label{eq:3-11c}\\
&  E_{22} = \frac{57}{98} Q_1 + \frac{51}{14} Q_2 + 3 Q_3
   - \frac14 Q_4 + \frac{16}{7} R_1 + \frac{30}{7} R_2,
\label{eq:3-11d}\\
&  E_{23} = - \frac37 Q_1 - 3 Q_2 + \frac12 Q_4 - \frac67 R_1,
\label{eq:3-11e}\\
&  E_{24} = \frac{3}{16} Q_1,
\label{eq:3-11f}\\
&  E_{33} = \frac37 Q_1 + \frac{27}{7} Q_2 + 2 Q_3 - \frac12 Q_4
   + \frac67 R_1 + \frac{12}{7} R_2,
\label{eq:3-11g}\\
&  E_{34} = -\frac38 Q_1 - \frac32 Q_2 + \frac14 Q_4,
\label{eq:3-11h}\\
&  E_{44} = \frac{3}{16} Q_1 + \frac32 Q_2 + \frac12 Q_3 - \frac14 Q_4,
\label{eq:3-11i}
\end{align}
and all the other $E_{nm}$ which are not listed above are zero. The
functions $Q_n(k)$ and $R_n(k)$ are given by
Eqs.~(\ref{eq:a-1a})--(\ref{eq:a-3b}) in Appendix~\ref{app:IntTens}.

As in the case of real space, the SPT result should be reproduced by
expanding the exponential prefactor in Eq.~(\ref{eq:3-9}). A formal
expression for the 1-loop power spectrum in redshift space
\cite{HMV98} contains 3-dimensional integrations. Further reduction of
the expression is possible and necessary to compare with our result.
In Appendix~\ref{app:RedPSSPT}, we derive a reduced expression for the
power spectrum of 1-loop SPT in redshift space, which itself is a
new result. Finally, the following relation, which is similar to
Eq.~(\ref{eq:2-1-1}), is confirmed:
\begin{align}
&  P_{\rm s}(\bm{k}) =
  \exp\left\{ - k^2[1 + f(f+2)\mu^2] A \right\}
\nonumber\\
& \quad \times
  \left\{ (1 + f\mu^2)^2 P_{\rm L}(k)
      + P_{\rm sSPT}^{\mbox{\scriptsize 1-loop}}(\bm{k})
  \right.
\nonumber\\
& \qquad\quad
  \left.
      +\, (1 + f\mu^2)^2[1 + f(f+2)\mu^2] k^2 P_{\rm L}(k) A
  \right\},
\label{eq:3-12}
\end{align}
where $P_{\rm sSPT}^{\mbox{\scriptsize 1-loop}}(\bm{k})$ is the 1-loop
contribution to the redshift-space power spectrum in SPT given in
Appendix~\ref{app:RedPSSPT}. Therefore, when the exponential prefactor
is expanded, Eq.~(\ref{eq:3-9}) reduces to the result of SPT.

The linear term in Eq.~(\ref{eq:3-9}) corresponds to the Kaiser
formula \cite{Kaiser87} of the linear perturbation theory. The
appearance of the exponential suppression prefactor is characteristic
in our formalism. This factor reflects the nonlinear smearing effect
of the velocity field along the lines of sight \cite{Jackson72}, such
as the fingers-of-God effect by large-scale random motions. The linear
theory and higher-order SPT do not have such a suppression factor
along the lines of sight.

This form of the suppression factor exactly matches a phenomenological
model of the redshift-space power spectrum on large scales
\cite{ESW07}, which is given by (after correcting obvious typos)
\begin{equation}
  P(\bm{k}) =
  \exp\left[
      -\frac12
      \left(
          k_{\perp}^2\sigma_{\perp}^2
          + k_{\parallel}^2\sigma_{\parallel}^2
      \right)
  \right]
  (1 + f\mu^2)^2 P_{\rm L}(\bm{k}),
\label{eq:3-13}
\end{equation}
where $k_\perp$ and $k_\parallel$ are wave vector components across
and along the line of sight, and $\sigma_\perp$ and $\sigma_\parallel$
are modeled by
\begin{equation}
  \sigma_\perp = s_0 D, \quad \sigma_\parallel = s_0 D (1 + f).
\label{eq:3-14}
\end{equation}
The length parameter $s_0$ characterizes the scale of radial
displacements. When all the 1-loop contributions of our
Eq.~(\ref{eq:3-9}) are omitted, the model of Eq.~(\ref{eq:3-13}) is
``derived'' if we identify $s_0 = (2A)^{1/2}/D = (2A_0)^{1/2}$, where
$A_0$ is a linear-theory value of $A$ at $z=0$. In Ref.~\cite{ESW07},
the parameter $s_0$ is measured from the $N$-body simulation of
Eisenstein, Seo, and White, and a value $s_0 = 9.40\, h^{-1}{\rm Mpc}$
is obtained for $100\,h^{-1}{\rm Mpc}$ separations in their cosmology
(we convert their quoted value according to different conventions of
the growth factor). Using the same set of cosmological parameters as
theirs, we find $A_0 = 44.0\, h^{-2}{\rm Mpc}^2$. The corresponding
prediction from our formalism is $s_0 = 9.38\, h^{-1}{\rm Mpc}$, which
is very close to their measured value.

There is another simple model of the nonlinear power spectrum in
redshift space \cite{Sco04}, which is given by
\begin{align}
&  P_{\rm s}(\bm{k})
   = \exp\left(-f^2 k^2 \mu^2 \sigma_v^2\right)
\nonumber\\
& \qquad\qquad \times
   \left[
       P_{\delta\delta}(k) + 2 f \mu^2 P_{\delta\theta}(k)
       + f^2 \mu^4 P_{\theta\theta}(k)
   \right],
\label{eq:3-15}
\end{align}
where $\sigma_v^2$ is the same as our $A$ of Eq.~(\ref{eq:3-10}), and
$P_{\delta\delta}(k)$, $P_{\delta\theta}(k)$, $P_{\theta\theta}(k)$
are (cross) power spectra of density and velocity fields, calculated
from nonlinear 1-loop SPT. Even though the density and velocity
(cross) correlations are taken into account, this model is no longer a
result of exact 1-loop perturbation theory (see
Appendix~\ref{app:RedPSSPT}). This simple model explains $N$-body data
to some extent but predicts a slightly larger power spectrum than
$N$-body data on small scales \cite{Sco04}. This model has a similar
structure to our result but lacks many terms.

The spherical average of Eq.~(\ref{eq:3-9}) is calculated by using the
following integral:
\begin{align}
&  \frac12 \int_{-1}^1 d\mu e^{-x \mu^2} \mu^{2n}
  = \frac12 x^{-n-1/2} \gamma\left(n+\frac12,x\right)
\nonumber\\
& \qquad\qquad\qquad
= (-1)^n \frac{\sqrt{\pi}}{2}
     \left(\frac{d}{dx}\right)^n
    \left[\frac{{\rm erf}(x^{1/2})}{x^{1/2}}\right],
\label{eq:3-20}
\end{align}
where $\gamma(a,x)$ is the lower incomplete gamma function, and ${\rm
  erf}$ is the error function normalized by ${\rm erf}(+\infty) = 1$.

\begin{figure*}
\begin{center}
\includegraphics[width=32pc]{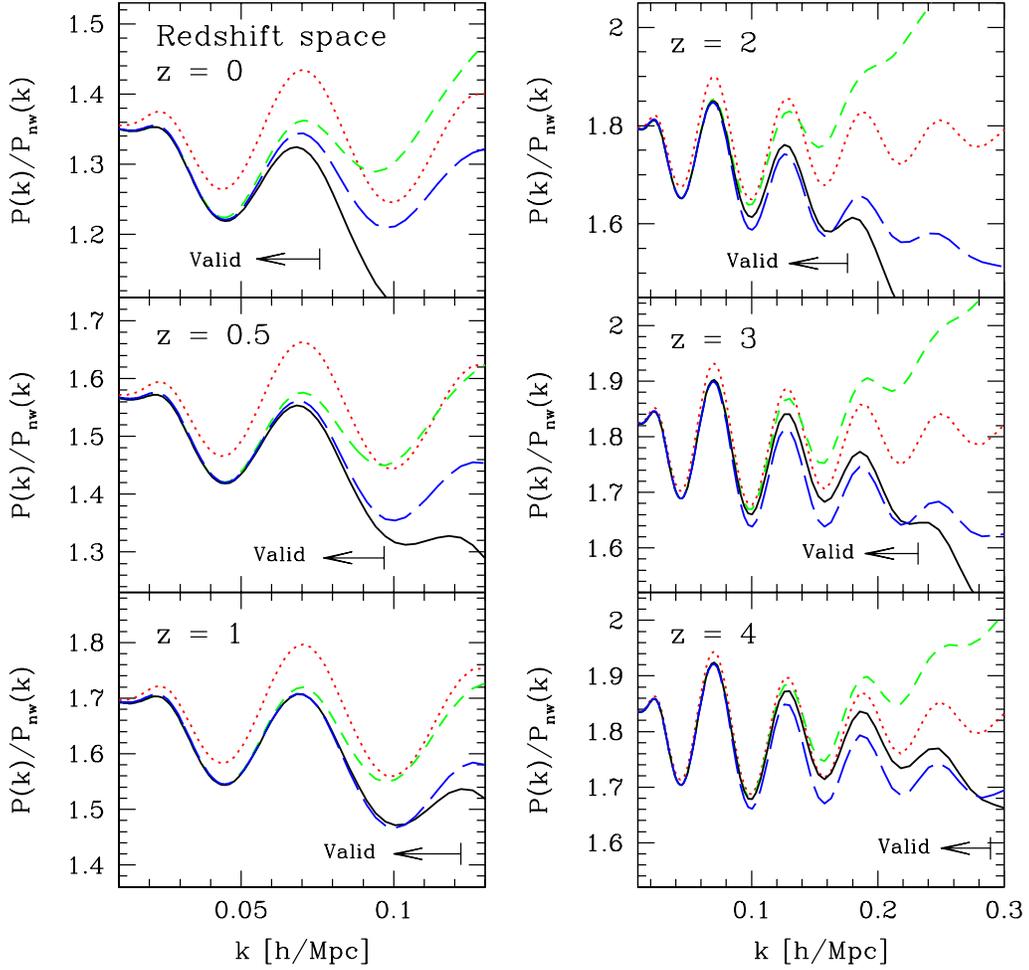}
\caption{\label{fig:redpsbao} Nonlinear evolution of the baryon
  acoustic oscillations in redshift space for various redshifts, $z =
  0$ (top left), $0.5$ (center left), $1$ (bottom left), $2$ (top
  right), $3$ (center right), $4$ (bottom right). Direction
  dependencies are spherically averaged. Each power spectrum is
  divided by a smoothed, no-wiggle linear power spectrum $P_{\rm
    nw}(k)$ in real space \cite{EH99}. {\em Black (solid) line}: this
  work; {\em red (dotted) line}: linear theory (Kaiser formula); {\em
    green (short dashed) line}: 1-loop SPT; {\em blue (long dashed)
    line}: a simple model of redshift-space power spectrum
  \cite{Sco04}. The validity range $k < k_{\rm NL}/2$, where our
  result is expected to be accurate within a few percent, is indicated
  by arrows.}
\end{center}
\end{figure*}

In Fig.~\ref{fig:redpsbao}, the spherically averaged nonlinear power
spectrum in redshift space is plotted on BAO scales.
The same cosmological parameters as in Fig.~\ref{fig:logps} are
adopted. Each power spectrum is normalized by a no-wiggle power
spectrum \cite{EH99} in real space. According to the Kaiser effect,
the amplitude is larger in higher redshifts. In the same plot, the
simple model of Eq.~(\ref{eq:3-15}) is overplotted. Both linear theory
and 1-loop SPT are larger than our result and the simple model, since
the former two do not have a suppression factor which corresponds to
nonlinear smearing effects along the lines of sight in redshift space.
The validity ranges $k < k_{\rm NL}/2$ are indicated in the figure,
where $k_{\rm NL}$ is given by Eq.~(\ref{eq:2-1-2}).

\begin{figure}
\begin{center}
\includegraphics[width=20pc]{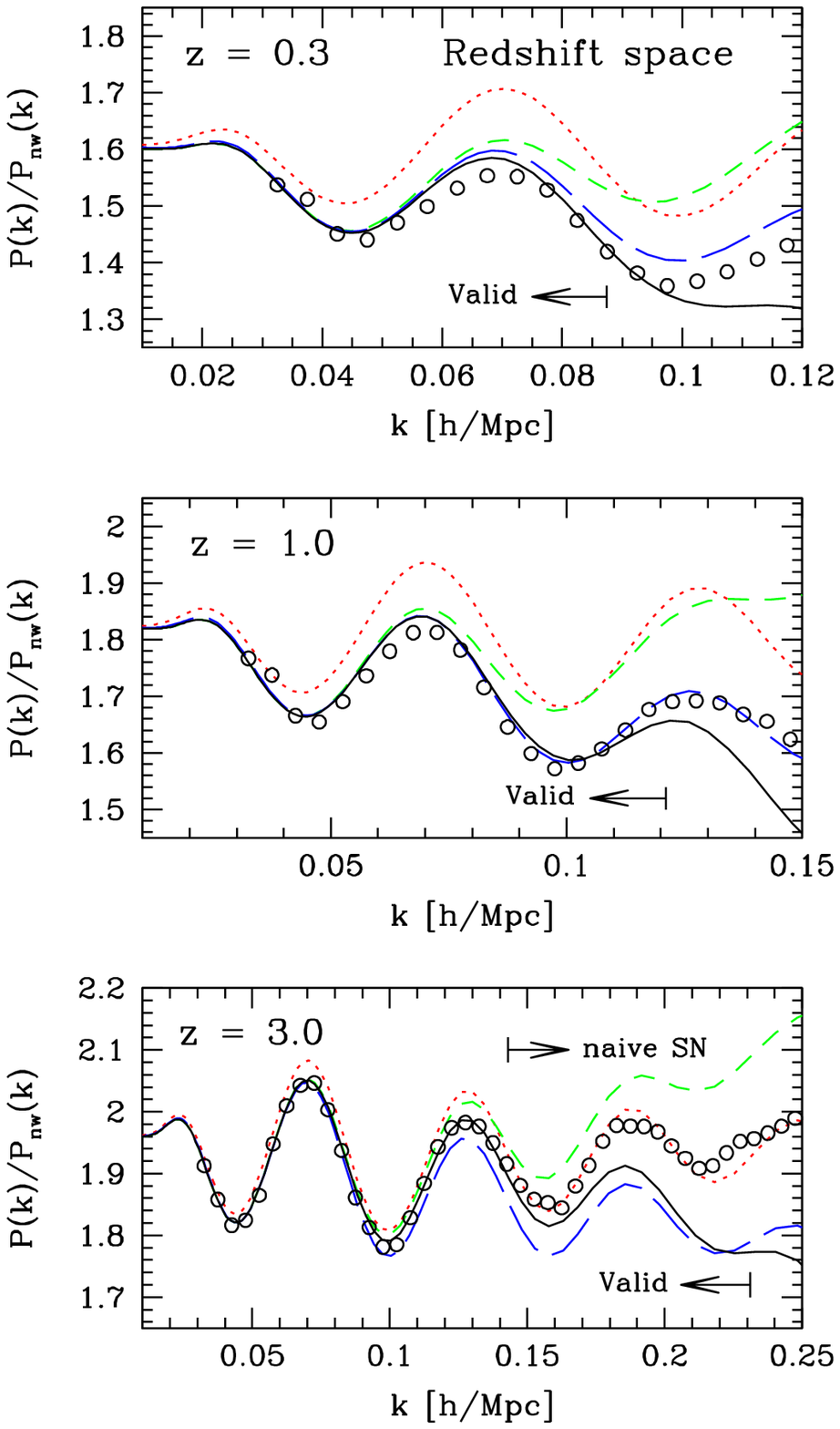}
\caption{\label{fig:psSEred} Comparison of the power spectra to the
  $N$-body simulations of Ref.~\cite{SE05} in redshift space at
  redshifts $z=0.3$, $1$, and $3$ from top to bottom, respectively.
  {\em Open circles}: $N$-body simulations; {\em Black (solid) line}:
  this work; {\em red (dotted) line}: linear theory; {\em green
    (dashed) line}: 1-loop SPT; {\em blue (long dashed) line}: a
  simple model of redshift-space power spectrum \cite{Sco04}. Only
  nonlinear deviations from the linear growth are measured in $N$-body
  simulations to reduce cosmic variances. Estimates of validity range,
  $k < k_{\rm NL}/2$ are indicated by arrows. In the $z=3$ sample, the
  range where contributions from the na\"ive shot noise exceed $1\%$
  is also indicated.}
\end{center}
\end{figure}

In Fig.~\ref{fig:psSEred}, a comparison of the power spectrum in
redshift space between our results and the $N$-body simulations of
Ref.~\cite{SE05} is presented.
As in real space, our results are better than 1-loop SPT results. The
1-loop SPT overestimates the power spectrum much worse in redshift
space. On smaller scales, the simple model of Eq.~(\ref{eq:3-15}) and
the linear theory seems to be good descriptions at a particular
redshift. However, that is just a coincidence because they do not
describe the simulations at different redshifts. The shot noise is not
corrected for the $N$-body simulation, and the $N$-body sample of
$z=3$ might be affected by the shot noise on small scales. The range
where the na\"ive shot noise is over $1\%$ is indicated in the figure.
Since the particle positions in redshift space are more randomly
distributed than in real space, the true shot noise might be larger
than that in real space. Overall, our results describe the $N$-body
results well at all redshifts.

\subsection{\label{subsec:CFRed}
The correlation function in redshift space
}

The correlation function $\xi_{\rm s}(\bm{r})$ in redshift space is
given by
\begin{equation}
  \xi_{\rm s}(\bm{r})
  = \int \frac{d^3k}{(2\pi)^3} e^{i\bm{k}\cdot\bm{r}}
  P_{\rm s}(\bm{k}).
\label{eq:3-21}
\end{equation}
When we consider the spherically averaged correlation function, the
relation of Eq.~(\ref{eq:2-2}) is applicable with the spherically
averaged power spectrum.

\begin{figure*}
\begin{center}
\includegraphics[width=32pc]{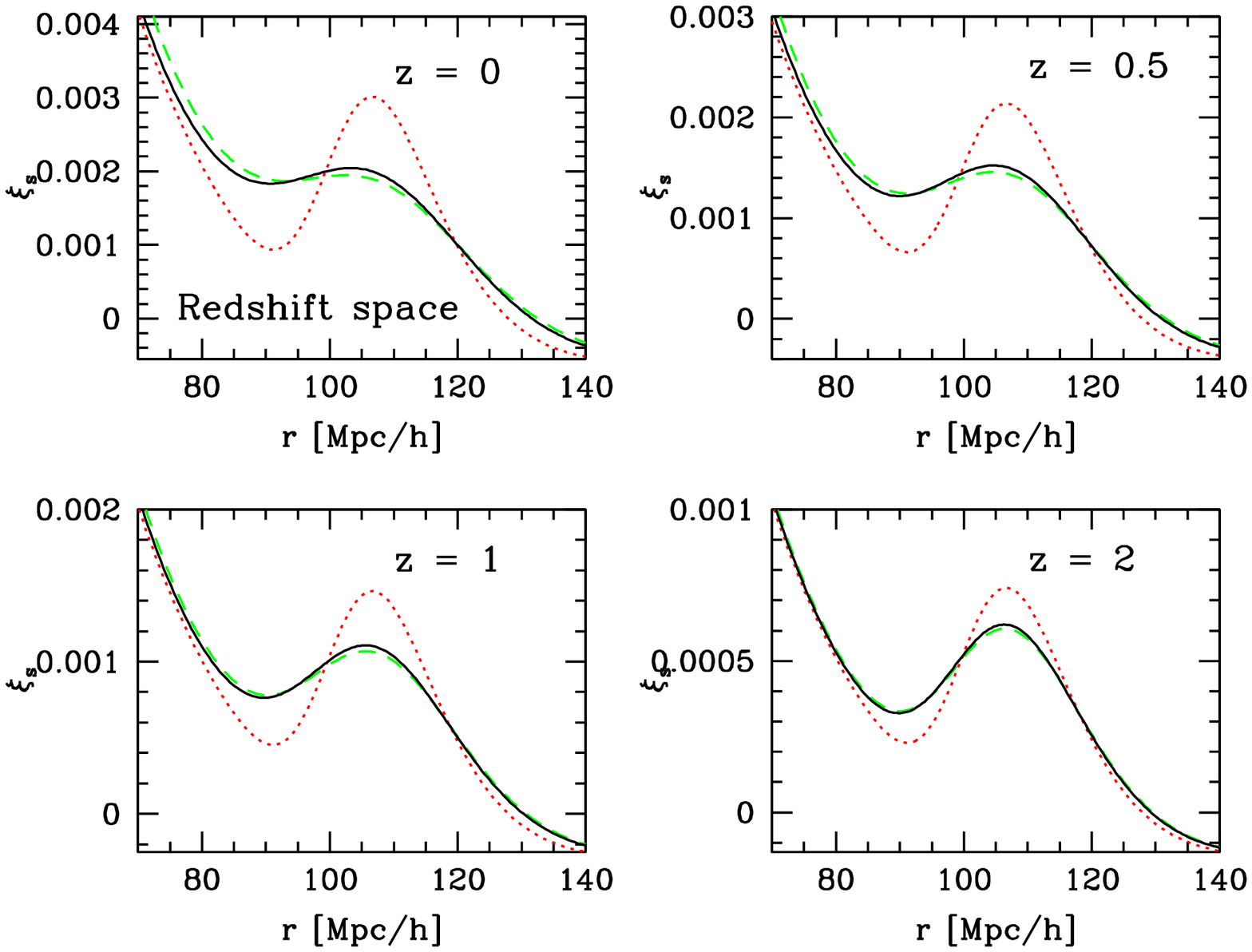}
\caption{\label{fig:redxibao} Nonlinear evolution of the baryon
  acoustic peak in redshift space for various redshifts, $z = 0$ (top
  left), $0.5$ (top right), $1$ (bottom left), $2$ (bottom right).
  Angular dependencies are averaged. {\em Black (solid) line}: this
  work; {\em red (dotted) line}: linear theory (Kaiser-Hamilton
  formula). {\em green (dashed) line}:
  Gaussian-filtered linear theory (see text).
}
\end{center}
\end{figure*}

In Fig.~\ref{fig:redxibao}, the spherically averaged nonlinear
correlation function in redshift space is plotted on BAO scales.
The same cosmological parameters as in Fig.~\ref{fig:realpsbao} are
adopted. Since the effect of the suppression factor is larger in
redshift space, the nonlinear smearing effects on the baryon peak are
larger than in real space (cf. Fig.~\ref{fig:realxibao}). This is
because the nonlinear redshift-space distortions additionally smear
the clustering along the line of sight. As in
Fig.~\ref{fig:realxibao}, Gaussian-filtered linear theory is
overplotted, which approximately explains the nonlinear smearing
effects.

\begin{figure}
\begin{center}
\includegraphics[width=19pc]{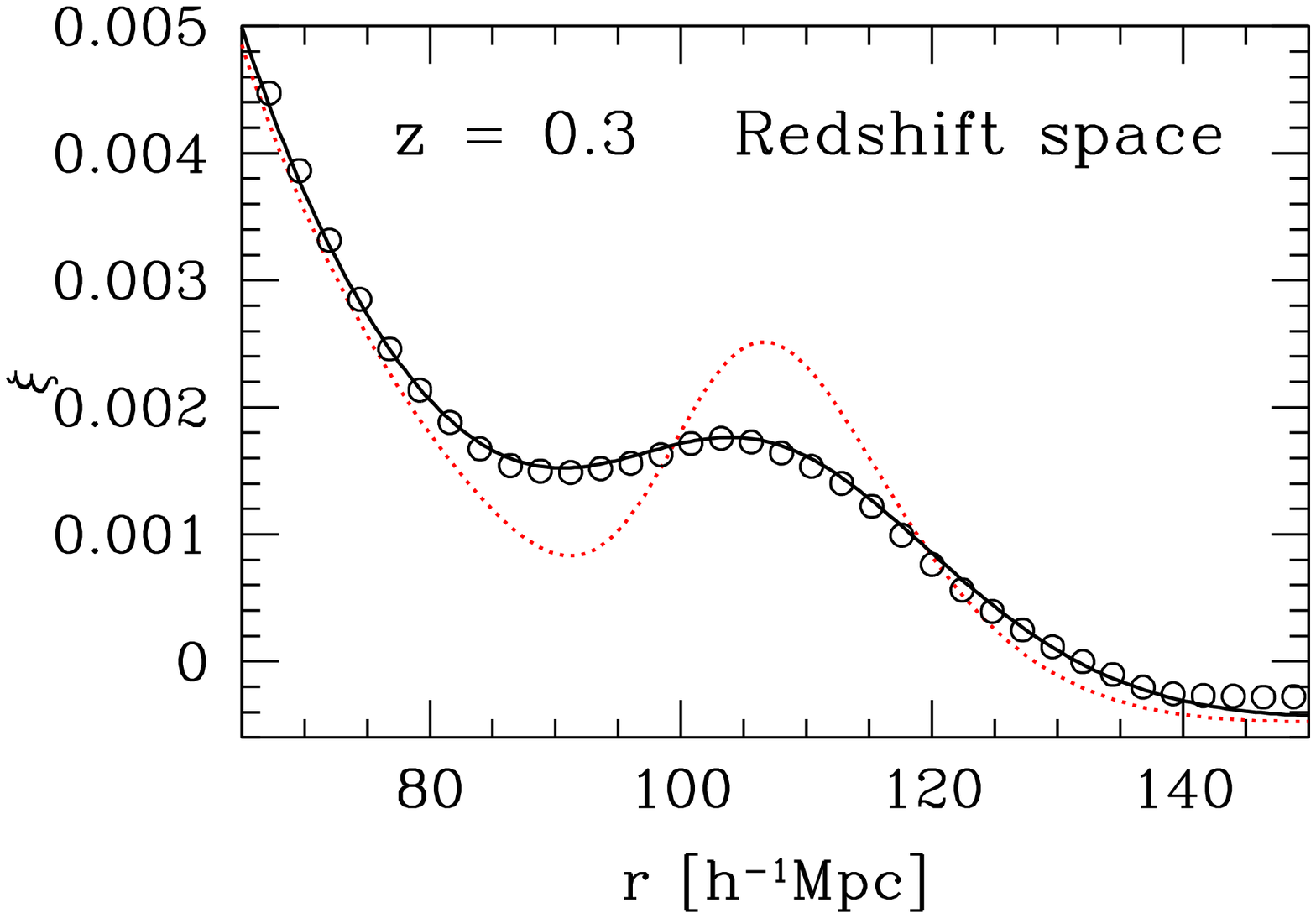}
\caption{\label{fig:xiSEred} Comparison of the correlation functions
  to the $N$-body simulations of Refs.~\cite{SE05,ESSS07} in redshift
  space. {\em Open circles}: $N$-body results; {\em Black (solid)
    line}: this work; {\em red (dotted) line}: linear theory. Only
  nonlinear deviations from the linear growth are measured in $N$-body
  simulations to reduce finiteness effects. }
\end{center}
\end{figure}

In Fig.~\ref{fig:xiSEred}, our result of the correlation function in
redshift space at $z=0.3$ is compared with the $N$-body simulations of
Refs.~\cite{SE05,ESSS07} using the same parameter as used in these
references.
The agreement is quite good also in redshift space, and our
result explains the nonlinear effects on BAO scales very precisely
even in redshift space.

\section{Conclusions
\label{sec:concl}
}

In this paper, we have formulated a new resummation method of the
cosmological perturbation theory, starting from the Lagrangian
description of perturbations. The infinite series of perturbations in
SPT is naturally contained in the exponential prefactor in our
formalism, and expanding the exponential factor reproduces the SPT. We
have presented 1-loop results of our formalism in this paper. Our
method is particularly useful for studying nonlinear effects on BAO
scales and always provides a better description of the $N$-body
results than SPT on large scales.

Nonlinear corrections to the correlation function can be calculated by
our formalism. This is a great advantage over SPT, since higher-order
SPT cannot predict the correlation function. Our 1-loop results are
comparable to the 1-loop results of the RPT \cite{CS07}. In the RPT,
it is shown that taking into account the 2-loop effects enlarges the
applicable range toward small scales in the power spectrum. Since our
formalism and RPT have common features in the resulting equations, we
expect that 2-loop extensions of our calculation similarly enlarge the
applicable range of scales in the power spectrum. The description of
the BAO peak in the correlation function seems already sufficient by
our 1-loop calculation.

It is an important and original outcome that our formalism gives an
analytical description of nonlinear effects in redshift space. The
large-scale structure is measured by redshift surveys of astronomical
objects, such as galaxies, quasars, Lyman-alpha absorbers, 21 cm
emission, etc. However, nonlinear corrections to the power spectrum
and correlation function in redshift space by perturbation theory have
not been sufficiently studied so far. Our formalism and results
presented in this paper are unprecedented in this respect and are
crucial steps toward applying the BAO measurements in redshift surveys
to constraining the nature of dark energy, the curvature of the
Universe, etc.

Considering the biasing effects in galaxy clustering is beyond the
scope of this paper. A simple prescription of the biasing is to
replace $f \rightarrow \beta = f/b$ in our equations, where $b$ is the
linear bias parameter. However, nonlinearity and nonlocality of
biasing might not be negligible even on BAO scales \cite{BAONL,
  ang07}. Incorporating the nonlinear and nonlocal biasing into our
formalism will be addressed in future work.

\begin{acknowledgments}
    I wish to thank H.-J.~Seo and D.~J.~Eisenstein for providing table
    forms of their figures in Refs.~\cite{SE05,ESSS07} and A.~Taruya
    and N.~Yoshida for discussion. I acknowledge support from the
    Ministry of Education, Culture, Sports, Science, and Technology,
    Grant-in-Aid for Scientific Research (C), 18540260, 2006, and
    Grant-in-Aid for Scientific Research on Priority Areas No. 467
    ``Probing the Dark Energy through an Extremely Wide and Deep
    Survey with Subaru Telescope.'' This work is supported in part by
    JSPS (Japan Society for Promotion of Science) Core-to-Core Program
    ``International Research Network for Dark Energy.''
\end{acknowledgments}

\appendix

\section{
\label{app:IntTens}
Integrating Cumulants of the Displacement Field}

In this appendix, we present results of integrating cumulants of the
displacement field, which appear in Eq.~(\ref{eq:1-27}). Although we
do not need to evaluate the integrations as tensors in real space, we
do need to do that in redshift space.

To present the results, it is convenient to define the following
integrals:
\begin{align}
&  Q_n(k) =  \frac{k^3}{4\pi^2}
  \int_0^\infty dr\, P_{\rm L}(kr) \int_{-1}^1 dx
\nonumber\\
& \qquad \quad \times
  P_{\rm L}[k(1 + r^2 - 2rx)^{1/2}]\,
  \frac{\tilde{Q}_n(r,x)}{(1 + r^2 - 2rx)^2},
\label{eq:a-1a}\\
&  R_n(k) =
   \frac{1}{48}
   P_{\rm L}(k) \frac{k^3}{4\pi^2}
    \int_0^\infty dr\, P_{\rm L}(kr)\,\tilde{R}_n(r),
\label{eq:a-1b}
\end{align}
where
\begin{align}
&  \tilde{Q}_1 = r^2 (1 - x^2)^2,
\label{eq:a-2a}\\
&  \tilde{Q}_2 = (1 - x^2) rx (1 - rx),
\label{eq:a-2b}\\
&  \tilde{Q}_3 = x^2 (1 - rx)^2,
\label{eq:a-2c}\\
&  \tilde{Q}_4 = 1 - x^2,
\label{eq:a-2d}
\end{align}
and
\begin{align}
&  \tilde{R}_1 = - \frac{2}{r^2} (1 + r^2)(3 - 14 r^2 + 3 r^4)
\nonumber\\
& \qquad\qquad\qquad\qquad
       + \frac{3}{r^3} (r^2 - 1)^4 \ln\left|\frac{1+r}{1-r}\right|,
\label{eq:a-3a}\\
&  \tilde{R}_2 = \frac{2}{r^2} (1 - r^2)(3 - 2 r^2 + 3 r^4)
\nonumber\\
& \qquad\qquad\qquad\quad
       + \frac{3}{r^3} (r^2 - 1)^3 (1 + r^2)
       \ln\left|\frac{1+r}{1-r}\right|.
\label{eq:a-3b}
\end{align}

The following expressions are directly obtained by substituting
Eqs.~(\ref{eq:1-22a})--(\ref{eq:1-23}) into
Eqs.~(\ref{eq:1-25a})--(\ref{eq:1-25c}):
\begin{align}
&  C^{(11)}_{ij}(\bm{k})
  = \frac{k_i k_j}{k^4} P_{\rm L}(k),
\label{eq:a-4a}\\
&  C^{(22)}_{ij}(\bm{k})
  = \frac{9}{98} \frac{k_i k_j}{k^4} Q_1(k),
\label{eq:a-4b}\\
&  C^{(13)}_{ij}(\bm{k}) =   C^{(31)}_{ij}(\bm{k})
  = \frac{5}{21} \frac{k_i k_j}{k^4} R_1(k).
\label{eq:a-4c}
\end{align}
In deriving Eq.~(\ref{eq:a-4c}), the transverse part of
Eq.~(\ref{eq:1-22c}) vanishes, because the rotational covariance
implies
\begin{equation}
  \int \frac{d^3p}{(2\pi)^3}
  g(\bm{k},\bm{p}) \bm{T}(\bm{k},-\bm{p},\bm{p}) \propto \bm{k},
\label{eq:a-5}
\end{equation}
where $g$ is a scalar function.

The remaining integrals are obtained by using the rotational
covariance. For example, when the integrand is given by $p_i p_j$
times a scalar function of $\bm{k}$ and $\bm{p}$, integration over
$\bm{p}$ should result in a form $X \delta_{ij} + Y k_i k_j$. The
coefficients $X$ and $Y$ can be obtained by contracting the original
expression with $\delta_{ij}$ and $k_i k_j$, and so forth. After these
kinds of manipulations, we obtain
\begin{align}
&  \int \frac{d^3p}{(2\pi)^3}
   C^{(112)}_{ijk}(\bm{k},-\bm{p},\bm{p}-\bm{k})
\nonumber\\
&\qquad
   = \int \frac{d^3p}{(2\pi)^3}
   C^{(121)}_{ijk}(\bm{k},-\bm{p},\bm{p}-\bm{k})
\nonumber\\
&\qquad
   = \frac{3}{14} \frac{k_i k_j k_k}{k^6} (R_1 + 2R_2)
    - \frac{3}{14} \frac{k_i \delta_{jk}}{k^4} R_1,
\label{eq:a-6a}\\
&  \int \frac{d^3p}{(2\pi)^3}
   C^{(211)}_{ijk}(\bm{k},-\bm{p},\bm{p}-\bm{k})
\nonumber\\
&\qquad
  = \frac{3}{14} \frac{k_i k_j k_k}{k^6} (Q_1 + 2Q_2)
    - \frac{3}{14} \frac{k_i \delta_{jk}}{k^4} Q_1,
\label{eq:a-6b}\\
&  \int \frac{d^3p}{(2\pi)^3}
   C^{(11)}_{(ij}(\bm{p}) C^{(11)}_{kl)}(\bm{k}-\bm{p})
\nonumber\\
&\qquad
  = \frac{3}{8}\frac{\delta_{(ij} \delta_{kl)}}{k^4} Q_1
  - \frac{1}{4}\frac{\delta_{(ij} k_k k_{l)}}{k^6}
     ( 3Q_1 + 12 Q_2 - 2 Q_4 )
\nonumber\\
&\qquad\quad
  + \frac{1}{8}\frac{k_{(i} k_j k_k k_{l)}}{k^8}
     (3Q_1 + 24Q_2 + 8Q_3 - 4Q_4),
\label{eq:a-6c}
\end{align}
where the spatial indices are symmetrized over in Eq.~(\ref{eq:a-6c}).

If we substitute those expressions of
Eqs.~(\ref{eq:a-4a})--(\ref{eq:a-4c}) and
(\ref{eq:a-6a})--(\ref{eq:a-6c}) into Eq.~(\ref{eq:1-27}), the
real-space result of Eq.~(\ref{eq:2-1}) is reproduced.



\section{
\label{app:RedPSSPT}
SPT calculation of the 1-loop power spectrum in redshift space}

In this appendix, the nonlinear power spectrum in redshift space is
calculated by 1-loop SPT. A formal expression found in the literature
\cite{HMV98, SCF99} is given by 3-dimensional integrals, which are
reduced to lower dimensional integrals below. It is suggested that the
1-loop calculation of the power spectrum in redshift space by SPT does
not give satisfactory results, because SPT in redshift space breaks
down at larger scales than in real space \cite{SCF99}. In this paper,
we do not recommend the extensive use of the 1-loop SPT in redshift
space. However, it is still useful to calculate the nonlinear
corrections by SPT in redshift space for the following reasons: First,
the 1-loop corrections indicate when the linear theory is applicable
and when nonlinear effects become important in redshift space. Second,
the 1-loop SPT gives the asymptotic behavior of nonlinear corrections
in the $k \rightarrow 0$ limit, which should be compared with any
nonlinear theory of redshift-space distortions on large scales.

The comoving real-space position $\bm{x}$ and the comoving
redshift-space position $\bm{s}$ are related by Eq.~(\ref{eq:3-2}), or
\begin{equation}
  \bm{s} = \bm{x} + \frac{v_z(\bm{x})}{aH}  \hat{\bm{z}}.
\label{eq:b-1}
\end{equation}
The density field in real space $\rho(\bm{x})$ and that in redshift
space $\rho_s(\bm{s})$ are related by conservation relation:
\begin{equation}
  \rho_{\rm s}(\bm{s}) d^3s = \rho(\bm{x}) d^3x,
\label{eq:b-2}
\end{equation}
Therefore, the density contrast in redshift space is given by
\begin{equation}
  \delta_{\rm s}(\bm{s}) = \left[1 + \delta(\bm{x})\right]J^{-1} - 1,
\label{eq:b-3}
\end{equation}
where $J = \partial(\bm{s})/\partial(\bm{x})$ is the Jacobian of the
mapping from real space to redshift space. One can easily calculate
the Fourier transform of Eq.~(\ref{eq:b-3}):
\begin{equation}
  \tilde{\delta}_{\rm s}(\bm{k})
  = \tilde{\delta}(\bm{k}) + 
  \int d^3x e^{-i\bm{k}\cdot\bm{x}}
  \left( e^{i k_z u_z} - 1 \right) \left[1 + \delta(\bm{x})\right],
\label{eq:b-4}
\end{equation}
where $u_z = - v_z/aH$. This relation is applicable even in the fully
nonlinear regime. In Fourier space, $\tilde{u}_z(\bm{k}) = i k_z
\tilde{\theta}(\bm{k})/k^2$.

Following the route of SPT, the density contrast and the peculiar
velocity, which are absent in an homogeneous universe, are expanded in
perturbative series:
\begin{eqnarray}
    \delta = \delta^{(1)} + \delta^{(2)} + \delta^{(3)} + \cdots ,
\label{eq:b-5a}\\
    \theta = \theta^{(1)} + \theta^{(2)} + \theta^{(3)} + \cdots ,
\label{eq:b-5b}
\end{eqnarray}
where $\theta = \bm{\nabla}\cdot\bm{v}/aH$ corresponds to the velocity
divergence. In the SPT, only growing-mode solutions in each order are
retained, and the peculiar velocity field is consistently assumed to
be irrotational. Thus, the velocity field is fully characterized by
the velocity divergence \cite{BCGS02}. In Fourier space, each
perturbative term is given by
\begin{align}
&    \tilde{\delta}^{(n)}(\bm{k}) =
    D^n \int\frac{d^3p_1}{(2\pi)^3}\cdots\frac{d^3p_n}{(2\pi)^3}
    (2\pi)^3\delta^3\left(\sum_{i=1}^n \bm{p}_i - \bm{k}\right)
\nonumber\\
&\qquad\qquad\qquad
    \times F_n(\bm{p}_1,\ldots,\bm{p}_n)
    \delta_0(\bm{p}_1) \cdots \delta_0(\bm{p}_n),
\label{eq:b-6a}\\
&    \tilde{\theta}^{(n)}(\bm{k}) =
    f D^n \int\frac{d^3p_1}{(2\pi)^3}\cdots\frac{d^3p_n}{(2\pi)^3}
    (2\pi)^3\delta^3\left(\sum_{i=1}^n \bm{p}_i - \bm{k}\right)
\nonumber\\
&\qquad\qquad\qquad
    \times G_n(\bm{p}_1,\ldots,\bm{p}_n)
    \delta_0(\bm{p}_1) \cdots \delta_0(\bm{p}_n).
\label{eq:b-6b}
\end{align}
The perturbative kernels $F_n$ and $G_n$ do not depend on time in an
Einstein--de~Sitter model. They weakly depend on time and cosmological
parameters in general cosmology, but it is still a good approximation
that the perturbative kernels are replaced by those in the
Einstein--de~Sitter model even in arbitrary cosmology \cite{BCGS02}.
The explicit form of the perturbation kernels is given in
Refs.~\cite{Fry84,JB94,Goroff86}.

Expanding the peculiar velocity field in Eq.~(\ref{eq:b-4}), and
applying the perturbative expansions of Eqs.~(\ref{eq:b-5a}) and
(\ref{eq:b-5b}), we obtain the perturbative series for the
redshift-space density field:
\begin{align}
&   \tilde{\delta}_{\rm s}(\bm{k}) =
    \sum_{n=1}^\infty D^n
    \int\frac{d^3p_1}{(2\pi)^3}\cdots\frac{d^3p_n}{(2\pi)^3}
    (2\pi)^3\delta^3\left(\sum_{i=1}^n \bm{p}_i - \bm{k}\right)
\nonumber\\
& \qquad\qquad\qquad
    \times S_n(\bm{p}_1,\ldots,\bm{p}_n)
    \delta_0(\bm{p}_1) \cdots \delta_0(\bm{p}_n),
\label{eq:b-7}
\end{align}
where the perturbative kernels in redshift space $S_n$ are given by
\begin{align}
    &S_1(\bm{p}_1) = 1 + f \mu^2,
\label{eq:b-8a}\\
    &S_2(\bm{p}_1,\bm{p}_2)
    = F_2(\bm{p}_1,\bm{p}_2) + f \mu^2 G_2(\bm{p}_1,\bm{p}_2)
\nonumber\\
    &\qquad
    + \frac12 f k \mu
    \left(\frac{p_{1z}}{p_1^2} + \frac{p_{2z}}{p_2^2}\right)
    + \frac12 (f k \mu)^2 \frac{p_{1z} p_{2z}}{p_1^2 p_2^2},
\label{eq:b-8b}\\
    &S_3(\bm{p}_1,\bm{p}_2,\bm{p}_3)
    = F_3(\bm{p}_1,\bm{p}_2,\bm{p}_3)
    + f \mu^2 G_3(\bm{p}_1,\bm{p}_2,\bm{p}_3)
\nonumber\\
    &\qquad
    + f k \mu \frac{p_{1z}}{p_1^2} F_2(\bm{p}_2,\bm{p}_3)
    + f k \mu \frac{p_{2z} + p_{3z}}{|\bm{p}_2 + \bm{p}_3|^2}
    G_2(\bm{p}_2,\bm{p}_3)
\nonumber\\
    &\qquad
    + (f k \mu)^2
    \frac{p_{1z} (p_{2z} + p_{3z})}{p_1^2|\bm{p}_2 + \bm{p}_3|^2}
    G_2(\bm{p}_2,\bm{p}_3)
\nonumber\\
    &\qquad
    + \frac12 (f k \mu)^2
    \frac{p_{1z} p_{2z}}{p_1^2 p_2^2}
    + \frac16 (f k \mu)^3 
    \frac{p_{1z} p_{2z} p_{3z}}{p_1^2 p_2^2 p_3^2},
\label{eq:b-8c}
\end{align}
up to third order, where $\mu = \bm{k}\cdot\hat{\bm{z}}/k$, $\bm{k} =
\bm{p}_1 + \cdots + \bm{p}_n$, and $p_{1z} =
\bm{p}_1\cdot\hat{\bm{z}}$ etc. The expression of $S_3$ should be
symmetrized over its arguments when necessary. These kernels are
equivalent to those derived in Ref.~\cite{SCF99}, while the derivation
and apparent expressions here are somewhat different from those of
Scoccimarro, Couchman, and Frieman. When the linear bias $b$ is
present, the perturbative kernels for density fluctuations of galaxies
are given by replacing $f \rightarrow \beta = f/b$ in the above
kernels. When the bias is modeled by a nonlinear and local one, it is
straightforward to include that by a method of Ref.~\cite{FG93}.

Given the perturbative kernels, the nonlinear power spectrum in
redshift space is calculated in a similar way to that in real space
\cite{MSS, JB94}. The power spectrum $P_{\rm s}(\bm{k})$ in redshift
space is defined by
\begin{equation}
  \left\langle
      \tilde{\delta}_{\rm s}(\bm{k}) \tilde{\delta}_{\rm s}(\bm{k}')
  \right\rangle_{\rm c} =
  (2\pi)^3 \delta^3(\bm{k} + \bm{k}') P_{\rm s}(\bm{k}).
\label{eq:b-9}
\end{equation}
Note that the power spectrum in redshift space is no longer isotropic.
The perturbative expansion of the power spectrum up to second order in
the linear power spectrum (i.e., fourth order in $\delta_0$) is given
by
\begin{equation}
    P_{\rm s}(\bm{k}) = D^2(t) P_{\rm s11}(\bm{k})
    + D^4(t) \left[P_{\rm s22}(\bm{k}) + P_{\rm s13}(\bm{k})\right],
\label{eq:b-10}
\end{equation}
where $P_{\rm s11}$ is the contribution from $\langle
\tilde{\delta}^{(1)} \tilde{\delta}^{(1)} \rangle$, $P_{\rm s22}$ is
from $\langle \tilde{\delta}^{(2)} \tilde{\delta}^{(2)} \rangle$, and
$P_{\rm s13}$ is from $\langle \tilde{\delta}^{(1)}
\tilde{\delta}^{(3)} \rangle + \langle \tilde{\delta}^{(3)}
\tilde{\delta}^{(1)} \rangle$. The first quantity is the linear (or
tree-level) power spectrum, and the last two quantities are 1-loop
corrections, which are given by
\begin{align}
    & P_{\rm s11}(\bm{k})
    = \left[ S_1(\bm{k}) \right]^2 P_0(k),
\label{eq:b-11a}\\
    & P_{\rm s22}(\bm{k})
    = 2 \int \frac{d^3p}{(2\pi)^3}
    \left[ S^{\rm s}_2(\bm{p},\bm{k}-\bm{p}) \right]^2
    P_0(p) P_0(|\bm{k}-\bm{p}|),
\label{eq:b-11b}\\
    & P_{\rm s13}(\bm{k})
    = 6 S_1(\bm{k}) P_0(k)
    \int \frac{d^3p}{(2\pi)^3}
    S^{\rm s}_3(\bm{k},\bm{p},-\bm{p})
    P_0(p),
\label{eq:b-11c}
\end{align}
where $S^{\rm s}_n$ is a symmetrized kernel obtained from $S_n$.
The first-order power spectrum is given by
\begin{equation}
    P_{\rm s11}(\bm{k})
    = (1 + f \mu^2)^2 P_0(k),
\label{eq:b-12}
\end{equation}
which is the Kaiser formula \cite{Kaiser87} in linear perturbation
theory.

Integration over the azimuthal angle can be analytically performed in
Eqs.~(\ref{eq:b-11b}) and (\ref{eq:b-11c}), and further integration
over the polar angle in Eq.~(\ref{eq:b-11c}) is also possible. Terms
which contain the line-of-sight components of the wave vector can be
evaluated by using the rotational covariance as described in
Appendix~\ref{app:IntTens}. After a lengthy but straightforward
calculation, the final results are given by
\begin{align}
&   P_{\rm s22}(\bm{k}) =
    \sum_{n,m} \mu^{2n} f^m
    \frac{k^3}{4\pi^2}
    \int_0^\infty dr P_0(kr)
    \int_{-1}^1 dx
\nonumber\\
    &\qquad \times
    P_0[k\,(1 + r^2 - 2rx)^{1/2}]
    \frac{A_{nm}(r,x)}{(1 + r^2 - 2rx)^2},
\label{eq:b-13a}\\
&   P_{\rm s13}(\bm{k}) =
    (1 + f \mu^2) P_0(k)
\nonumber\\
    &\qquad \times
    \sum_{n,m} \mu^{2n} f^m
    \frac{k^3}{4\pi^2}
    \int_0^\infty dr P_0(kr) B_{nm}(r)
\label{eq:b-13b}
\end{align}
where non-vanishing components of $A_{nm}$ and $B_{nm}$ are
\begin{align}
&   A_{00} = \frac{1}{98} (3r + 7x - 10 rx^2)^2,
\label{eq:b-14a}\\
&   A_{11} = 4 A_{00},
\label{eq:b-14b}\\
&   A_{12} = \frac{1}{28} (1 - x^2)(7 - 6r^2 - 42rx + 48r^2x^2),
\label{eq:b-14c}\\
&   A_{22} = \frac{1}{196}
    \left[
        - 49 + 637 x^2 + 42rx(17 - 45x^2)
    \right.
\nonumber\\ &\qquad\qquad\qquad\qquad
    \left.
        + 6r^2(19 -157x^2 + 236x^4)
    \right],
\label{eq:b-14d}\\
&   A_{23} = \frac{1}{14} (1 - x^2)(7 - 42rx - 6r^2 + 48 r^2 x^2),
\label{eq:b-14e}\\
&   A_{24} = \frac{3}{16} r^2 (1 - x^2)^2,
\label{eq:b-14f}\\
&   A_{33} = \frac{1}{14}
    (-7 + 35x^2 + 54rx - 110rx^3
\nonumber\\ &\qquad\qquad\qquad\qquad
 + 6r^2 - 66r^2x^2 + 88r^2x^4),
\label{eq:b-14g}\\
&   A_{34} = \frac18 (1 - x^2)(2 - 3r^2 - 12rx + 15r^2x^2),
\label{eq:b-14h}\\
&   A_{44} = \frac{1}{16}
    (-4 + 12x^2 + 3 r^2 + 24rx
\nonumber\\ &\qquad\qquad\qquad\qquad
 - 30r^2x^2 - 40rx^3 + 35r^2x^4),
\label{eq:b-14i}
\end{align}
and
\begin{align}
&   B_{00} = \frac{1}{252} \left[
    \frac{12}{r^2} - 158 + 100 r^2 - 42 r^4
    \right.
\nonumber\\ &\qquad\qquad\quad
    \left.
    + \frac{3}{r^3}
    (r^2 - 1)^3(7r^2 + 2) \ln\left|\frac{1+r}{1-r}\right|
    \right],
\label{eq:b-15a}\\
&   B_{11} = 3 B_{00},
\label{eq:b-15b}\\
&   B_{12} = \frac{1}{168} \left[
    \frac{18}{r^2} - 178 - 66 r^2 + 18 r^4
    \right.
\nonumber\\ &\qquad\qquad\qquad
    \left.
    - \frac{9}{r^3}
    (r^2 - 1)^4 \ln\left|\frac{1+r}{1-r}\right|
    \right],
\label{eq:b-15c}\\
&   B_{22} = \frac{1}{168} \left[
    \frac{18}{r^2} - 218 + 126 r^2 - 54 r^4
    \right.
\nonumber\\ &\qquad\qquad\quad
    \left.
    + \frac{9}{r^3}
    (r^2 - 1)^3(3r^2 + 1) \ln\left|\frac{1+r}{1-r}\right|
    \right],
\label{eq:b-15d}\\
&   B_{23} = - \frac23.
\end{align}
In a limit $f \rightarrow 0$, only $m=n=0$ terms survive, and the
previous results of the 1-loop power spectrum in real space \cite{MSS}
are recovered.




\newcommand{\apjl}{Astrophys. J. Letters}
\newcommand{\apjs}{Astrophys. J. Suppl. Ser.}
\newcommand{\mnras}{Mon. Not. R. Astron. Soc.}
\newcommand{\pasj}{Publ. Astron. Soc. Japan}
\newcommand{\apss}{Astrophys. Space Sci.}
\newcommand{\aap}{Astron. Astrophys.}
\newcommand{\physrep}{Phys. Rep.}
\newcommand{\mpla}{Mod. Phys. Lett. A}
\newcommand{\jcap}{J. Cosmol. Astropart. Phys.}


\end{document}